\documentclass[preprint,12pt]{aastex}

\newcommand{\ee}[1]{\mbox{${} \times 10^{#1}$}}

\newcommand{\am}{\mbox{\arcmin}}
\newcommand{\as}{\mbox{\arcsec}}

\newcommand{\um}{$\mu$m}

\newcommand{\lsun}{\mbox{L$_\odot$}}
\newcommand{\msun}{\mbox{M$_\odot$}}

\newcommand{\nthp}{N$_2$H$^+$}
\input{epsf}

\newcommand{\andre}{Andr\'{e}}

\slugcomment{\footnotesize {}}
\shorttitle{SHARC-II Observations of Cores}
\shortauthors{Wu et al.}
\begin{document}
\title {\bf SHARC-II mapping of Spitzer c2d Small Clouds and Cores}
\author{
Jingwen Wu\altaffilmark{1,2},
Michael M. Dunham\altaffilmark{1,3},
Neal J. Evans II\altaffilmark{1},
Tyler L. Bourke\altaffilmark{4},
and Chadwick H. Young\altaffilmark{5}}

\altaffiltext{1}{Department of Astronomy, The University of Texas at Austin, 1 University Station, C1400, Austin, Texas 78712--0259}

\altaffiltext{2}{E-mail: jingwen@astro.as.utexas.edu}

\altaffiltext{3}{E-mail: mdunham@astro.as.utexas.edu}

\altaffiltext{4}{Harvard-Smithsonian Center for Astrophysics, 60 Garden Street, Cambridge, MA 02138}

\altaffiltext{5}{Department of Physical Sciences, Nicholls State University, Thibodaux, LA 70301}

\begin{abstract}

We present the results of a submillimeter survey of 53 low-mass dense cores with the Submillimeter High Angular Resolution Camera II (SHARC-II).  The survey is a follow-up project to the \emph{Spitzer} Legacy Program ``From Molecular Cores to Planet-Forming Disks'', with the purpose being to create a complete data set of nearby low-mass dense cores from the infrared to the millimeter.  We present maps of 52 cores at 350 $\mu$m and three cores at 450 $\mu$m, two of which were observed at both wavelengths. Of these 52 cores, 41 were detected by SHARC-II:  32 contained one submillimeter source while 9 contained multiple sources.  For each submillimeter source detected, we report various source properties including source position, fluxes in various apertures, size, aspect ratio, and position angle.  For the 12 cores that were not detected we present upper limits.  The sources detected by SHARC-II have, on average, smaller sizes at the 2$\sigma$ contours than those derived from longer-wavelength bolometer observations.  We conclude that this is not caused by a failure to integrate long enough to detect the full extent of the core; instead it arises primarily from the fact that the observations presented in this survey are insensitive to smoothly varying extended emission.  We find that SHARC-II observations of low-mass cores are much better suited to distinguishing between starless and protostellar cores than observations at longer wavelengths.  Very Low Luminosity Objects, a new class of objects being discovered by the \emph{Spitzer Space Telescope} in cores previously classified as starless, look very similar at 350 \um\ to other cores with more luminous protostars.
\end{abstract}



\section{Introduction}\label{intro}

Many questions remain concerning the formation of low-mass stars despite several decades of study.  One of the primary reasons is that stars form in extremely dense regions of dust and gas, obscuring nearly all the light emitted in the optical portion of the spectrum.  In the near- and mid-infrared the opacities of the dust grains are low enough that light can begin to escape.  Thus, images at these wavelengths can reveal the central object (a protostar and possibly a circumstellar disk).  The \emph{Spitzer Space Telescope} Legacy project ``From Molecular Cores to Planet-Forming Disks'' (c2d; Evans et al. 2003) has carried out a survey of low-mass dense cores with the Infrared Array Camera (IRAC; Fazio et al. 2004) and the Multiband Imaging Photometer (MIPS; Rieke et al. 2004), providing images ranging from 3.6 to 70 $\mu$m.  \emph{Spitzer} is much more sensitive than previous space infrared missions, providing an ability to detect fainter, less-luminous sources than was possible with previous infrared surveys.  In fact, \emph{Spitzer} c2d observations have shown that several cores believed to be starless (containing no protostars) actually harbor Very Low Luminosity Objects (VeLLOs) with internal luminosities\footnote{The internal luminosity of an object, $L_{int}$, is defined to be the luminosity of the central object (star and disk, if present), and excludes luminosity arising from heating by the Interstellar Radiation Field (ISRF) (Di Francesco et al. 2006).} $L_{int} \leq 0.1$ \lsun\ (e.g. Young et al. 2004,  Di Francesco et al. 2006).  Infrared images can reveal a wealth of information about the internal sources embedded in dense cores, but the infrared alone does not tell the whole story.

To fully understand the process of low-mass star formation and the nature of embedded protostars, one must understand the details of the dusty envelopes in which they are embedded.  Even though these envelopes are usually quite cold ($\sim$ 10 K; e.g. Di Francesco et al. 2006), except for the inner regions where they are heated by the internal source, their thermal emission can easily be detected at submillimeter and millimeter wavelengths.  Complementary surveys are obtaining continuum observations at 850 $\mu$m with the Submillimetre Common User's Bolometer Array (SCUBA) at the James Clerk Maxwell Telescope (JCMT) (Young et al. 2006), at 1.2 mm with with the Max Planck Millimeter Bolometer (MAMBO) at the IRAM 30 meter telescope (J. Kauffmann et al. 2006, in preparation), and also at 1.2 mm with the SEST Imaging Bolometer Array (SIMBA) at the Swedish ESO Submillimeter Telescope (SEST) (K. Brede et al. 2006, in preparation).  SCUBA also provides observations at 450 \um, although often of much lower quality.  However, since the peak of a 15 K blackbody occurs at 340 $\mu$m, a reliable measurement that is close to this wavelength and fills the gap between 70 $\mu$m (\emph{Spitzer}) and 850 $\mu$m (SCUBA) is essential to constrain the peak of the Spectral Energy Distributions (SEDs) of these objects.  The MIPS instrument aboard \emph{Spitzer} can also provide images at 160 $\mu$m, but such data are unavailable for many of the low-mass cores of interest.  Thus, SHARC-II, with its capability to provide high-quality data at 350 $\mu$m, is ideally suited to fill this gap.

SHARC-II is a background-limited 350 and 450 $\mu$m facility camera mounted on the Caltech Submillimeter Observatory (CSO) at Mauna Kea, Hawaii (Dowell et al. 2002). It adopts the advanced ``CCD-style'' bolometer array with 12 $\times$ 32 pixels, resulting in a 2\farcm59 $\times$ 0\farcm97 field of view. SHARC-II features relatively high angular resolution: it has a beam size of 8\farcs5 at 350 $\mu$m and $\sim$11$\arcsec$ at 450 $\mu$m with good focus and pointing.  Since the atmospheric transmission is very sensitive to the weather at the higher frequencies at which SHARC-II operates, it only works well in very dry weather.  Under optimal weather conditions ($\tau_{225\ GHz} < 0.05$), SHARC-II can reach an RMS noise of 25 mJy beam$^{-1}$ in one hour of integration at 350 $\mu$m.

Because SHARC-II achieves optimum performance at 350 $\mu$m, a wavelength at which high-quality data is often difficult to obtain, and because the peak of a 15 K blackbody occurs closer to 350 $\mu$m than 450 $\mu$m, we mapped our sources at 350 $\mu$m.  We also present three sources that were mapped at 450 $\mu$m.  In this paper, we present the basic data of this survey, including images and source properties, in a manner that is consistent with other papers presenting submillimeter and millimeter continuum data on the c2d cores and clouds.  We leave a detailed analysis combining all available data on these objects to later papers.  \S \ref{obs} discusses the observations, \S \ref{reduction} details the data reduction, \S \ref{results} discusses the preliminary results of this survey, \S \ref{starless} discusses the differences between SHARC-II maps of cores with and without protostars embedded within them, and \S \ref{conclusions} presents our conclusions.

\section{Observations}\label{obs}

\subsection{Target Selection}\label{targets}

The first column of Table \ref{info} lists each core observed, in order of Right Ascension.  This survey is designed to provide a complementary dataset to the c2d observations of low-mass dense cores; thus most of our targets are selected from the list of cores observed by that project.  For the earlier observation runs that occurred before the \emph{Spitzer} data was available, we chose what we believed to be the most scientifically interesting cores from the Molecular Cloud Cores Database\footnote{http://cfa-www.harvard.edu/sirtf-c2d-cfa/CORES-DB/}, created by members of the c2d team at the Harvard-Smithsonian Center of Astrophysics (CfA).  However, not all of these cores made the final \emph{Spitzer} target list, so some of the cores in our sample do not have corresponding \emph{Spitzer} data.  Once the \emph{Spitzer} data became available, we focused on cores found to be interesting and worthy of follow-up submillimeter observations.  In addition to the observations of dense cores, c2d also obtained maps of five large molecular clouds and IRS data on a number of sources, and some of our targets are drawn from these two projects.  Finally, some targets are chosen from cores observed in both a number of follow-up \emph{Spitzer} General Observer (GO) programs and by other observers with Guaranteed Time Observations (GTO) data.

Consequently, this survey is certainly not unbiased.  The primary criteria for selecting sources from the various projects were as follows: (1) provide complementary data to the \emph{Spitzer} c2d data, (2) create as much overlap as possible between the various programs obtaining complementary data at submillimeter and millimeter wavelengths, and (3) obtain submillimeter observations of the most scientifically interesting cores.  Regarding the second criterion, Table \ref{overlap} lists whether or not each core observed in this survey is included in surveys with SCUBA (850 \um; Young et al. 2006), MAMBO (1.2 mm; J. Kauffmann et al. 2006, in preparation), and SIMBA (1.2 mm; K. Brede et al. 2006, in preparation), as well as a complementary program to obtain $\sim$1\am\ resolution \nthp\ (1-0) and CS (2-1) maps with the Five College Radio Astronomy Observatory (FCRAO; C. De Vries et al. 2006, in preparation) and a project to construct radiative transfer models of starless cores using 450 and 850 \um\ data from the SCUBA archive (Y. Shirley et al. 2006, in preparation).  This is by no means a complete list of projects in which continuum and molecular line data on these cores can be found, especially for the cores in Perseus since many recent studies have focused on that region (e.g., Enoch et al. 2006; Kirk et al. 2006); it is only meant to compare the cores observed in this survey with other, complementary programs currently in progress.

As of November 2005, we have mapped 52 cores at 350 $\mu$m and three at 450 $\mu$m, two of which were observed at both wavelengths.  Out of the 53 unique cores observed here, 18 are included in the SCUBA survey, 16 are included in the MAMBO survey, 6 are included in the SIMBA survey, 18 are included in the FCRAO survey, and 7 are included in the radiative transfer modeling project by Y. Shirley et al. (2006, in preparation).

\subsection{Observations}\label{subobs}

Observations were conducted in May and September of 2003, June and September of 2004, and March, June and November of 2005 at the CSO.  We used the sweep mode of SHARC-II without chopping to observe all our sources. In this mode the telescope moves in a Lissajous pattern that keeps the central regions of the maps fully sampled.  Beyond this region of uniform coverage, the integration time per position decreases smoothly from the center outward.  According to information presented on the SHARC-II webpage\footnote{http://www.submm.caltech.edu/\~\ sharc/}, this mode works best for sources with sizes less than or comparable to the size of the array.  During all of our observations except those obtained in June 2005, the Dish Surface Optimization System (DSOS\footnote{See http://www.cso.caltech.edu/dsos/DSOS\_MLeong.html}) was used to correct the dish surface figure for gravitational deformations as the dish moves in elevation during observations.  For the June 2005 observations, an equipment malfunction caused the DSOS to correct all observations for a zenith angle of $49\mbox{$^{\circ}$}$ (where the zenith angle is defined to be $90\mbox{$^{\circ}$}-\mathrm{elevation}$), regardless of the actual zenith angle of the observations.  The effects of this on our results are discussed in \S \ref{results}.

Table \ref{info} lists, for each core observed in this survey, the coordinates of the center of the map, the adopted distance to the core, the project from which the core was selected, the date(s) the core was observed, and the 1$\sigma$ RMS noise of the map, in units of both mJy beam$^{-1}$ and \msun\ beam$^{-1}$. The coordinates of the map centers were taken from either the positions of interesting objects in the \emph{Spitzer} observations (if available), the location of molecular emission peaks from literature (e.g., Lee et al. 2001; Caselli et al. 2002), or positions suggested by collaborators to optimize the mapping coverage.  The B59 core consists of a cluster of low mass sources, and in this survey we only observed the one that is most likely associated with B59-MMS1 (Reipurth et al. 1996; Brooke et al. 2006).  Integrations on each source were separated into blocks of ten minutes in stable weather or five minutes in unstable (variable) weather.  The pointing was checked every 1-2 hours each night, primarily with planets such as Mars, Uranus, and Neptune.  If no planets were available we used secondary objects such as CRL618, IRC+10216, and IRAS 16293-2422.  After averaging over all the runs, the blind pointing uncertainty is 2\farcs1 for azimuth and 3\farcs1 for zenith angle. But since we corrected the pointing after each check, these actually represent upper limits.  For any given observation, the pointing uncertainty should be smaller than this. The pointing sources were also used as flux calibrators (see Section \ref{cal}).  The RMS noise of each map was calculated based on the statistics of the off-source regions of the maps.

\section{Data Reduction and Calibration}\label{reduction}

\subsection{Data Reduction}\label{subreduction}

All of the raw scans were reduced with the Comprehensive Reduction Utility for SHARC-II (CRUSH), a publicly available\footnote{http://www.submm.caltech.edu/\~\ sharc/crush/index.htm}, Java-based software package.  CRUSH iteratively solves a series of models that attempt to reproduce the observations, taking into account both instrumental and atmospheric effects (e.g., Beelen et al. 2006).  To first order, sky emission is subtracted by removing emission common to all bolometers (see Enoch et al. 2006 for a discussion of this process in relation to the Bolocam bolometer array at the CSO).  Observations obtained through the end of 2004 were reduced with version 1.35 of CRUSH, while observations taken in 2005 were reduced with version 1.40a9-2.  We tested the two versions by reducing several sources with both version 1.35 and version 1.40a9-2 and found the results to be consistent between the two versions of the reduction software.

As mentioned above, the Lissajous scan pattern results in a map with less integration time per pixel as the distance from the center of the map increases.  Thus, the map is better sampled in the center than at the edges, resulting in noise at the edges that can be mistaken for real emission.  To compensate for this, we used ``imagetool,'' a tool available as part of the CRUSH package, to eliminate the regions of each map that had a total integration time less than 25\% of the maximum.  This eliminates most, but not all, of the spurious edge emission.

\begin{figure}[hbt!]
\epsscale{0.90}
\plotone{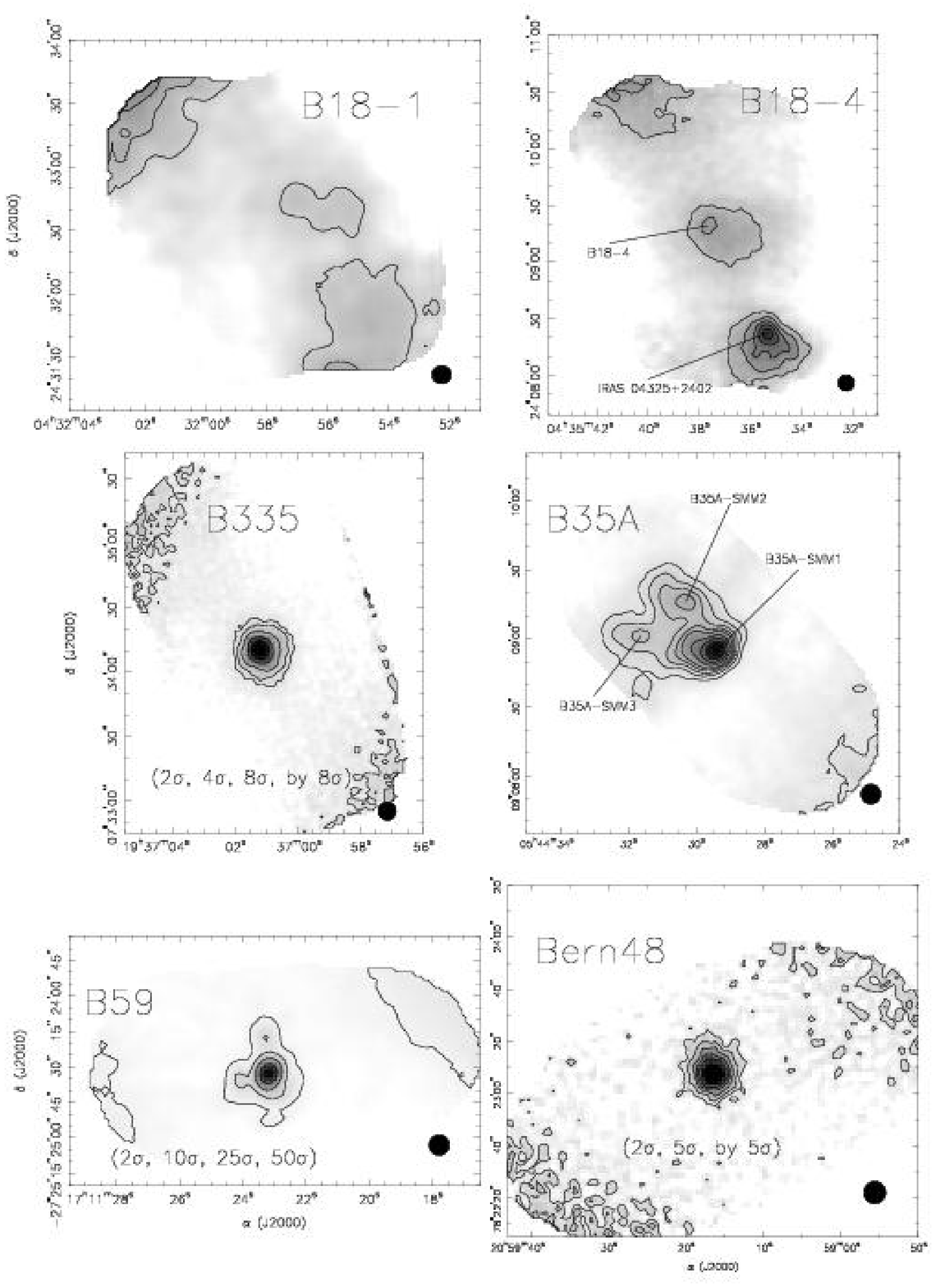}
\caption{\label{f1}SHARC-II 350 $\mu$m maps of the cores listed in in Table \ref{info}. Contours begin at 2$\sigma$ and increase by 2$\sigma$, unless otherwise indicated.  The beam size is shown at the lower right of each map, and cores with multiple sources have each source labeled.  Emission seen towards the edges of the maps is not reliable and should be ignored.}
\end{figure}

\begin{figure}[hbt!]
\plotone{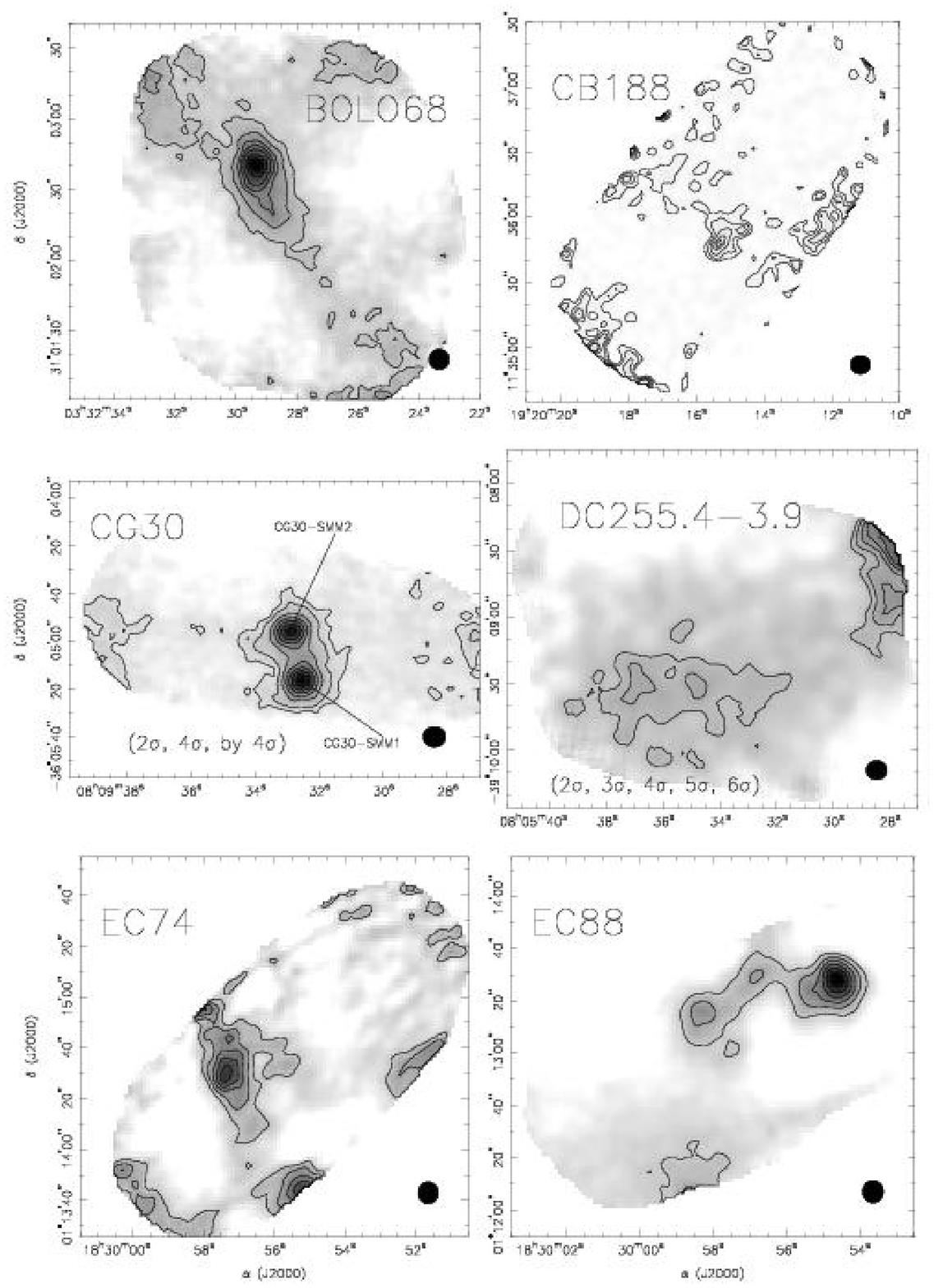}
\caption{\label{f2}SHARC-II 350 $\mu$m maps, continued.  Emission seen towards the edges of the maps is not reliable and should be ignored.}
\end{figure}

\begin{figure}[hbt!]
\plotone{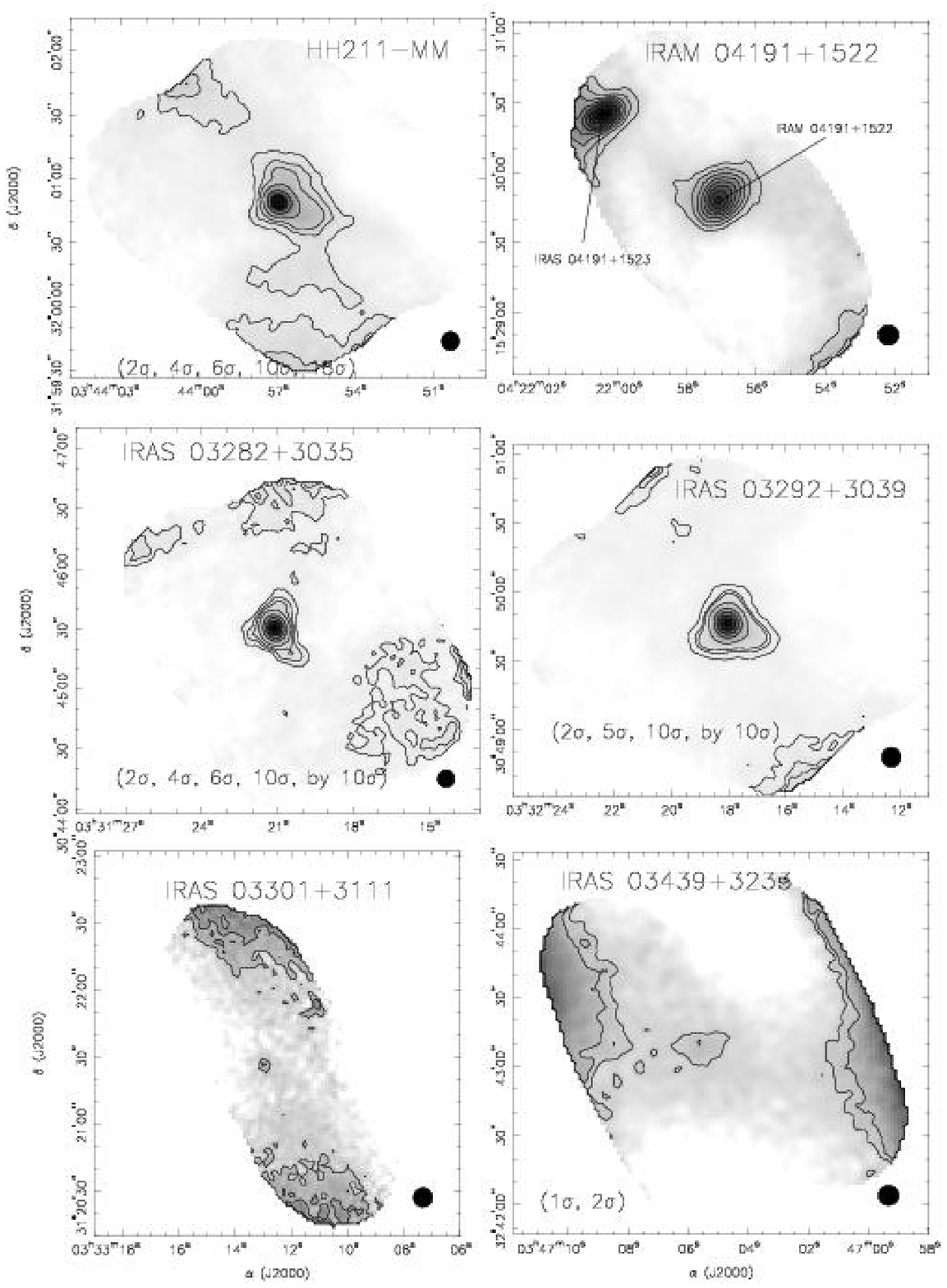}
\caption{\label{f3}SHARC-II 350 $\mu$m maps, continued.  Emission seen towards the edges of the maps is not reliable and should be ignored.}
\end{figure}

\begin{figure}[hbt!]
\plotone{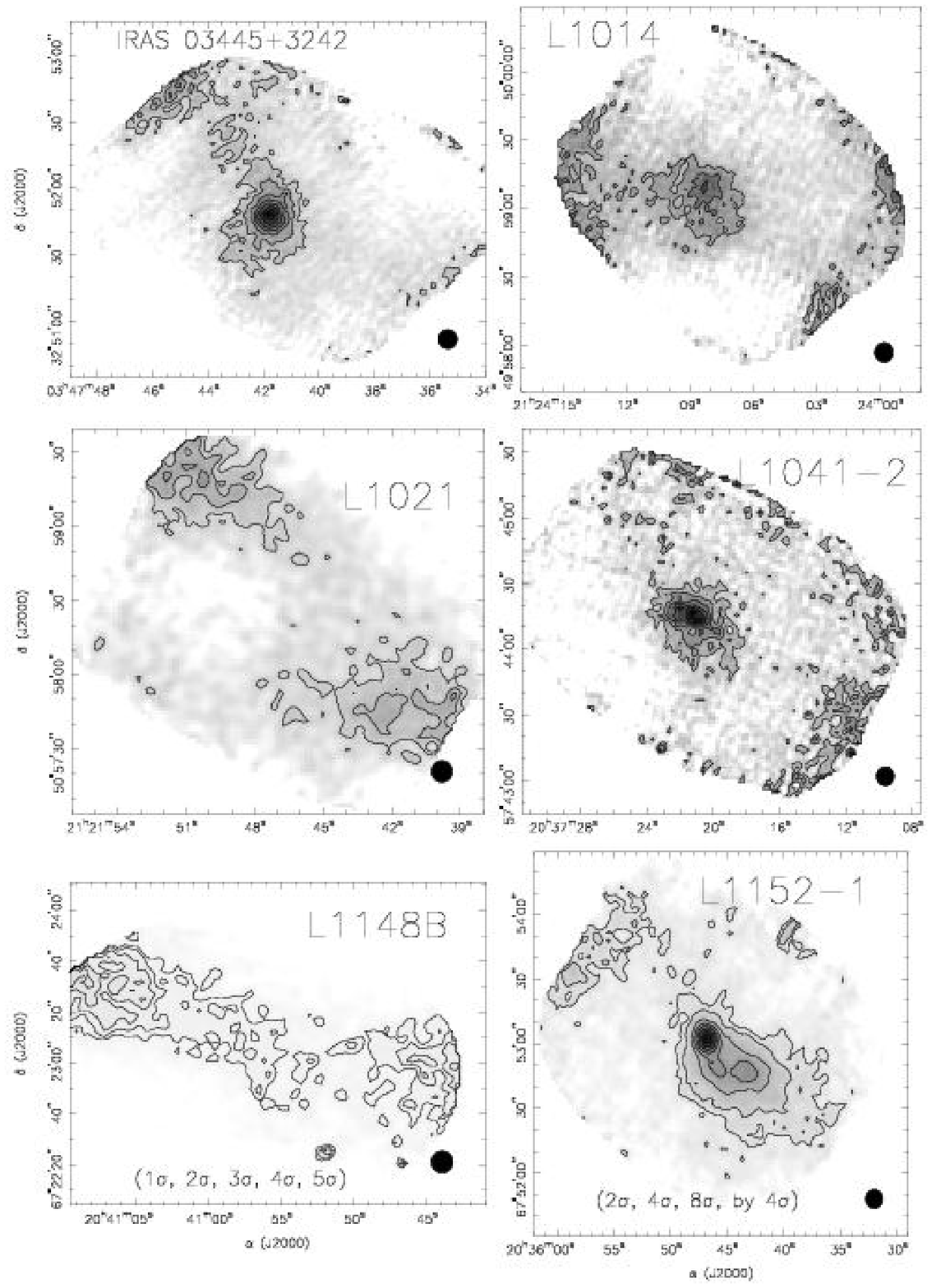}
\caption{\label{f4}SHARC-II 350 $\mu$m maps, continued.  Emission seen towards the edges of the maps is not reliable and should be ignored.}
\end{figure}

\begin{figure}[hbt!]
\plotone{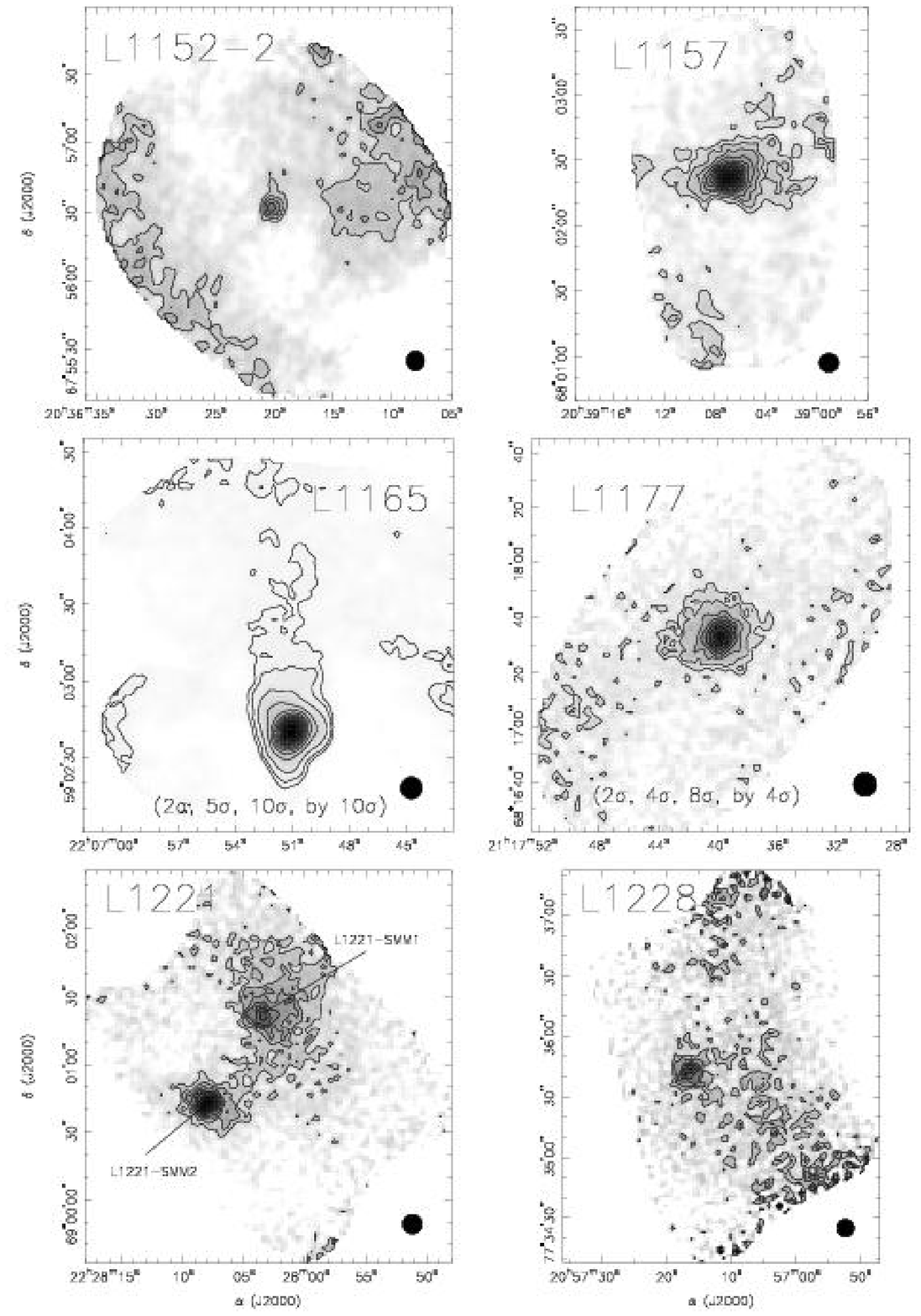}
\caption{\label{f5}SHARC-II 350 $\mu$m maps, continued.  Emission seen towards the edges of the maps is not reliable and should be ignored.}
\end{figure}

\begin{figure}[hbt!]
\plotone{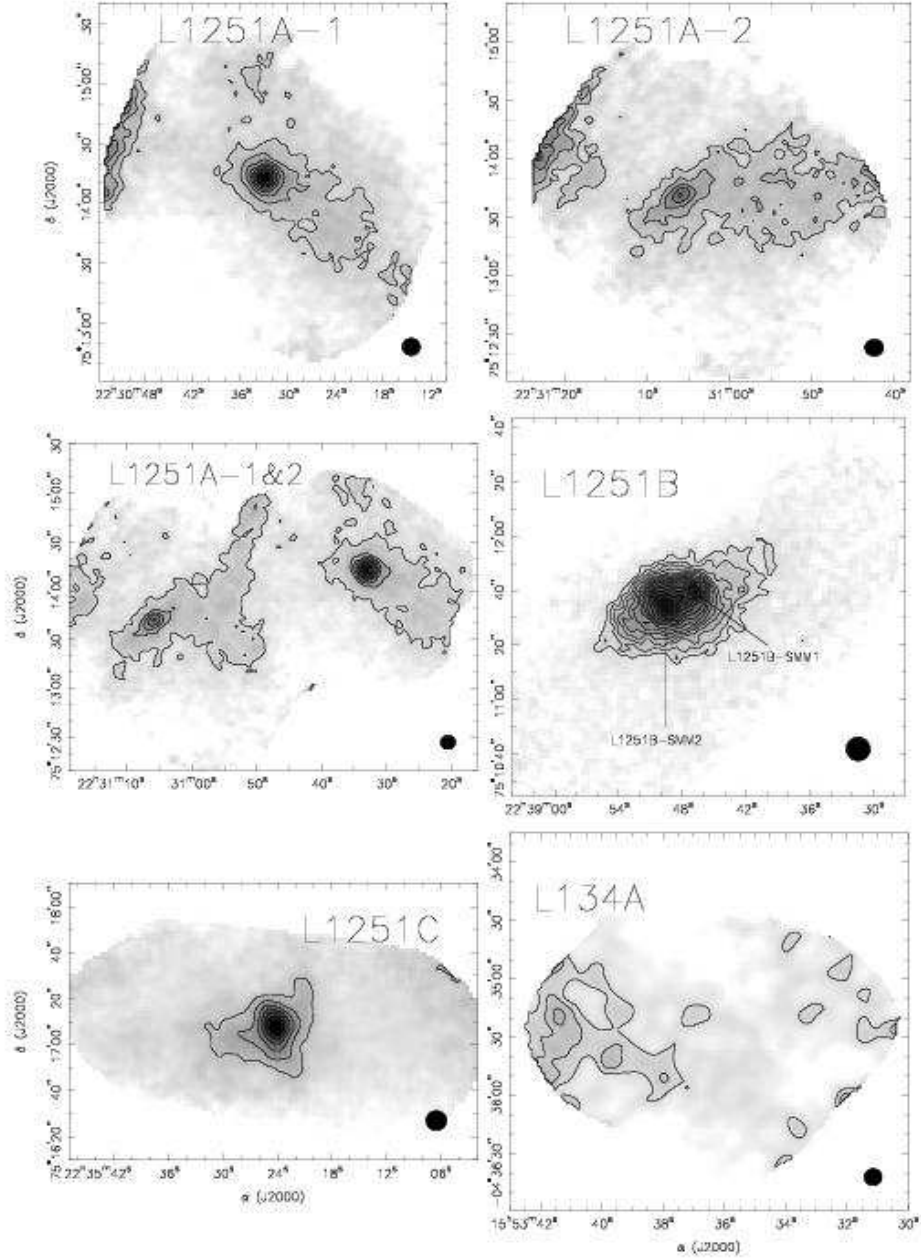}
\caption{\label{f6}SHARC-II 350 $\mu$m maps, continued.  Emission seen towards the edges of the maps is not reliable and should be ignored.  The combined map of L1251A-1 and L1251A-2 includes a filament between the two sources extending south from the northern edge of the map.  This results from spurious edge emission at the edges of the individual maps of L1251A-1 and L1251A-2 and should not be treated as real emission.}
\end{figure}

\begin{figure}[hbt!]
\plotone{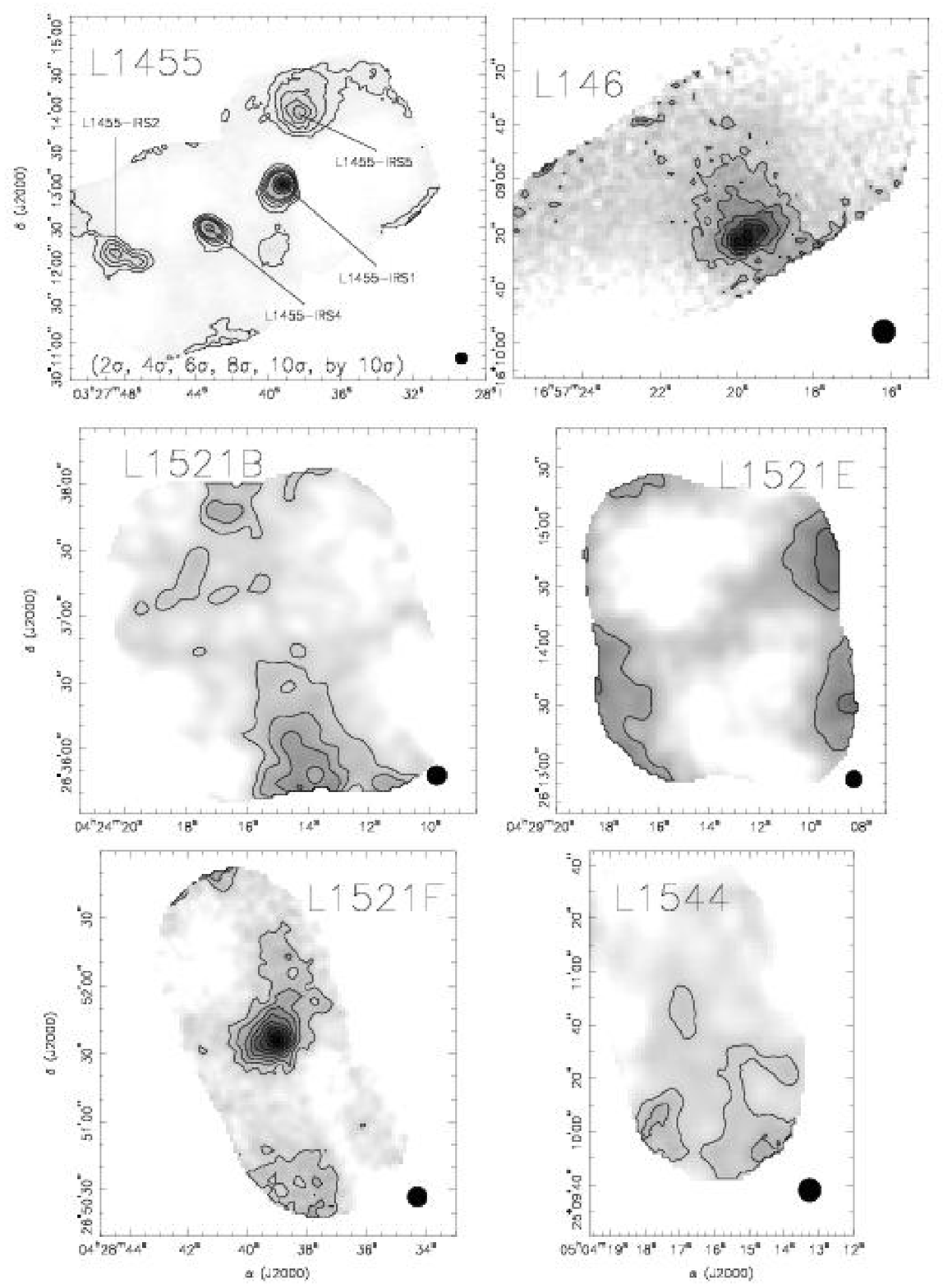}
\caption{\label{f7}SHARC-II 350 $\mu$m maps, continued.  Emission seen towards the edges of the maps is not reliable and should be ignored.}
\end{figure}

\begin{figure}[hbt!]
\plotone{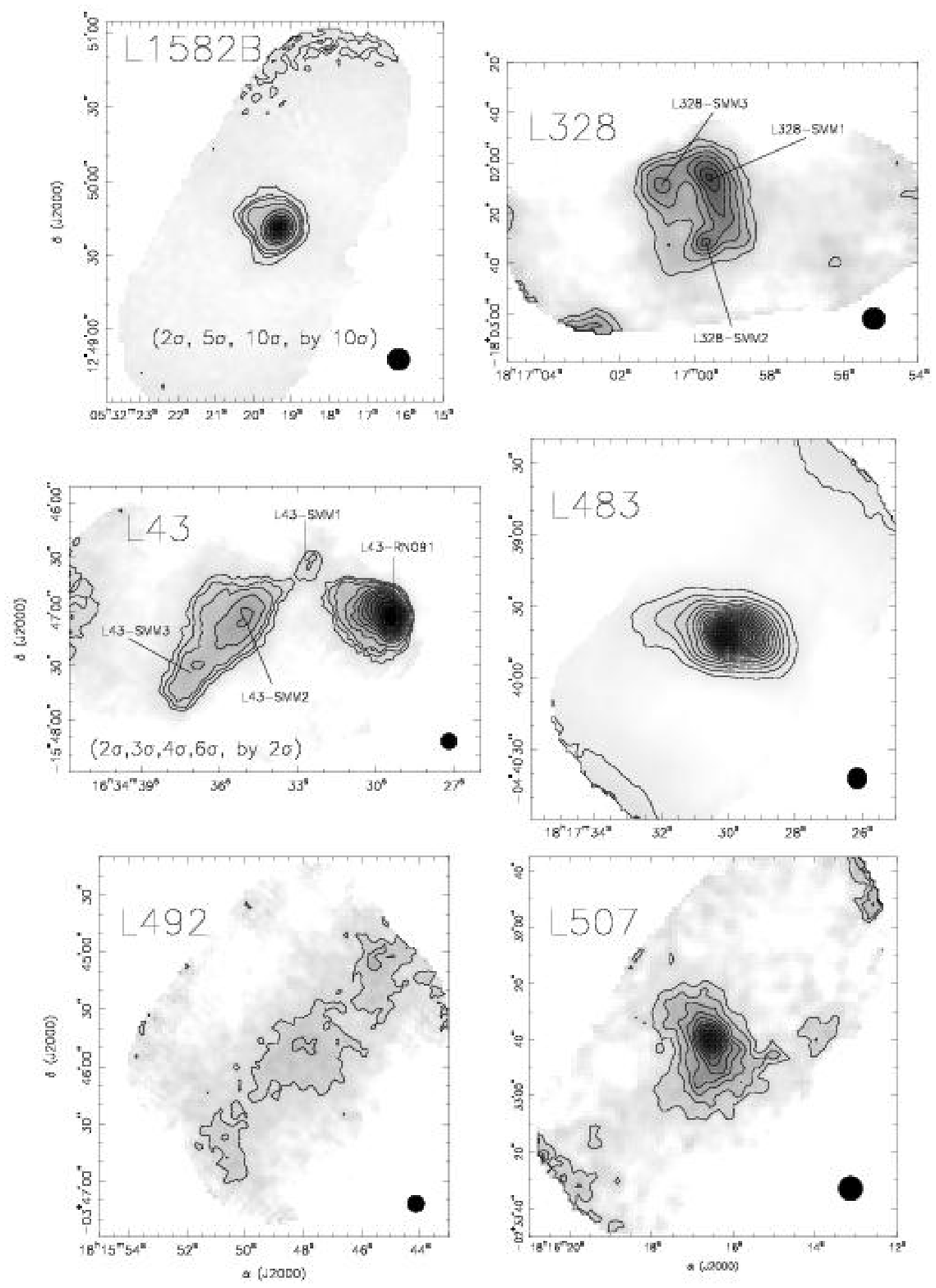}
\caption{\label{f8}SHARC-II 350 $\mu$m maps, continued.  Emission seen towards the edges of the maps is not reliable and should be ignored.}
\end{figure}

\begin{figure}[hbt!]
\plotone{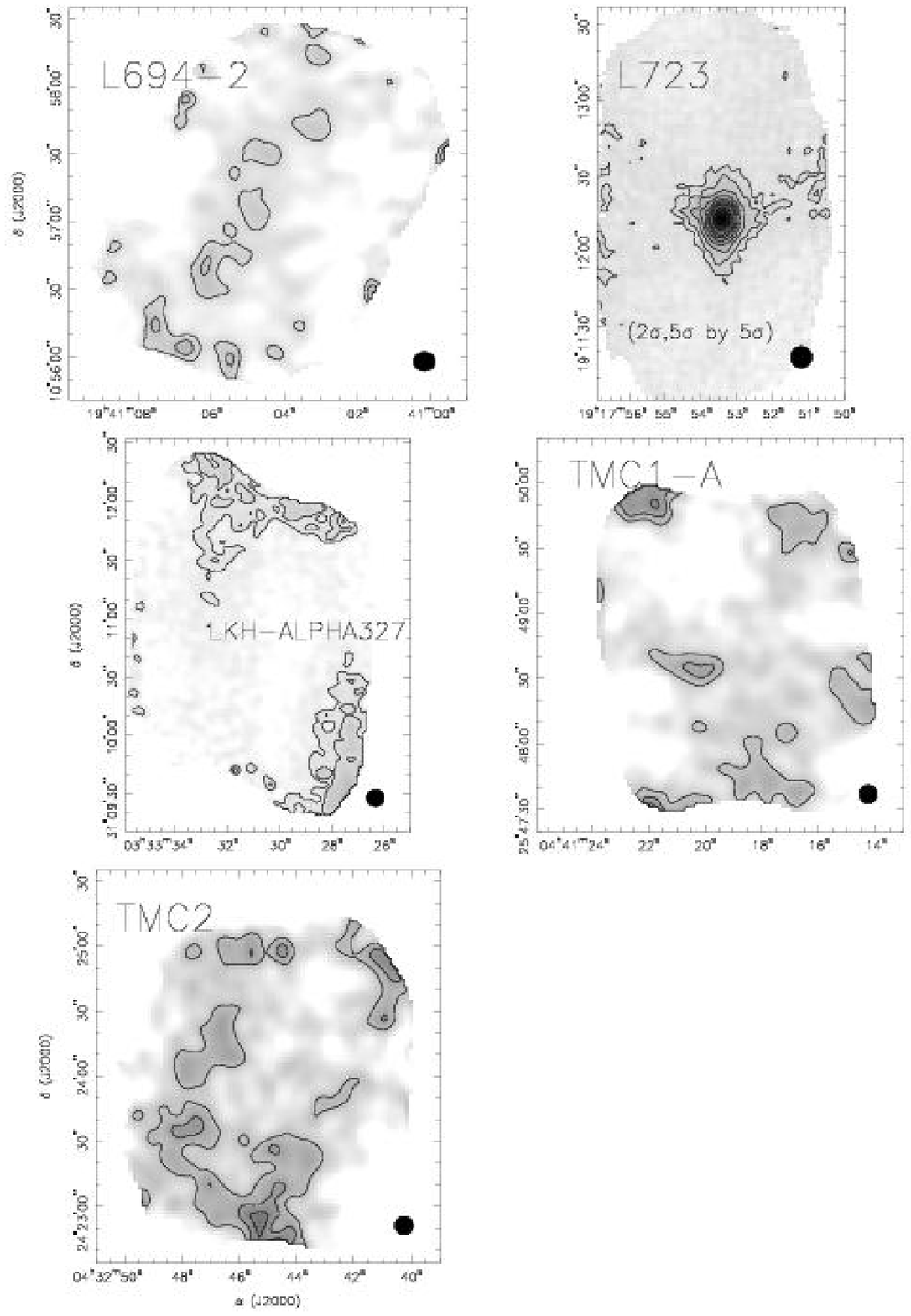}
\caption{\label{f9}SHARC-II 350 $\mu$m maps, continued.  Emission seen towards the edges of the maps is not reliable and should be ignored.}
\end{figure}

\begin{figure}[hbt!]
\plotone{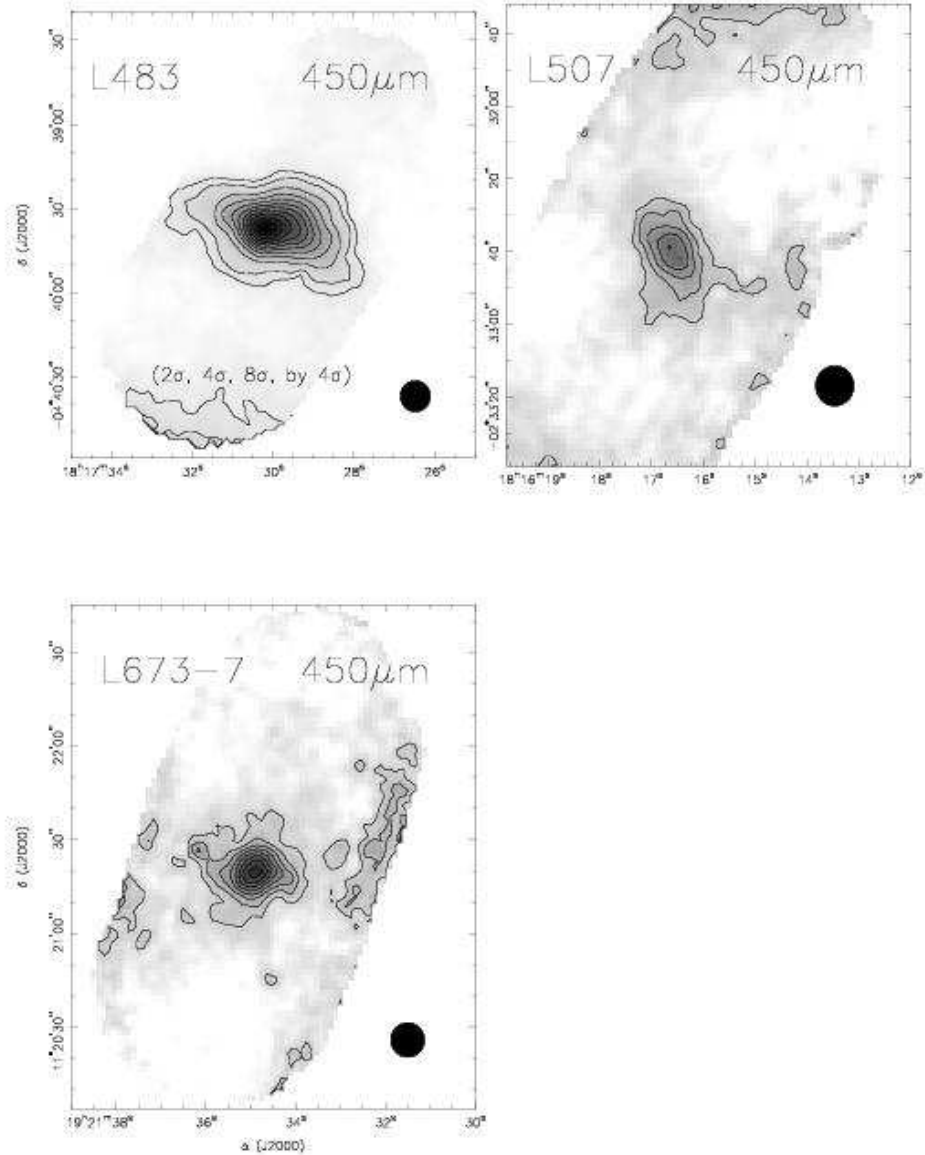}
\caption{\label{f450}SHARC-II 450 $\mu$m maps of the three cores listed in Table \ref{info} observed at this wavelength. Contours begin at 2$\sigma$ and increase by 2$\sigma$, unless otherwise indicated.  The beam size is shown at the lower right of each map.  Emission seen towards the edges of the maps is not reliable and should be ignored.}
\end{figure}

After removing the poorly sampled map edges, we used Starlink's ``stats'' package to assess the 1$\sigma$ RMS noise of the map, calculated using all of the pixels in the off-source regions.  Because this calculation includes all off-source pixels with total integration time greater than 25\% of the maximum, the RMS noise can be biased by the pixels with total integration times only slightly greater than 25\% of the maximum.  To investigate this, we used the feature of GAIA\footnote{http://star-www.rl.ac.uk/star/dvi/sun214.htx/sun214.html} that allows for the calculation of statistics in any user-defined region to assess the RMS noise of off-source pixels in the most well-sampled regions possible.  We conclude that the 1$\sigma$ RMS in the fully-sampled, central regions of the maps may be up to a factor of two lower than the values presented in Table \ref{info}, but because the integration time (and thus the noise) varies smoothly in these maps, the values given in Table \ref{info} represent the best estimate of the noise over the full map.  Finally, we used the derived RMS of each map to create contour maps overlaid on greyscale images for each core, and we present these in Figures \ref{f1}-\ref{f9} for the cores observed at 350 \um\ and in Figure \ref{f450} for those observed at 450 \um.  Contours start at $2\sigma$ and increase by $2\sigma$ unless otherwise indicated.  We arrange the images in alphabetical order rather than by Right Ascension to make a specific map easier to find.

\subsection{Calibration}\label{cal}

To measure the flux densities of the sources in a given aperture in units of Jy (as opposed to the instrument units of nV), we have calculated Flux Conversion Factors (FCF) for each aperture.  A brief description of the calibration method, explained in greater detail in Shirley et al. (2000), is as follows.  The FCF for an aperture of diameter $\theta$, C$_{\theta}$, is defined to be the total flux density of a calibrator source (a source with a known flux density at the wavelength of the observations) in Jy divided by the flux density in the same aperture of that calibrator in the instrument units.  Since CRUSH includes an atmospheric correction, the flux density of a source in a given aperture is then obtained by simply multiplying the flux of the source in that aperture, in the instrument units, by the FCF for that aperture.  A set of standard apertures is adopted for all of the papers presenting complementary millimeter and submillimeter data to the c2d project:  20\arcsec, 40\arcsec, 80\arcsec, and 120\arcsec.  Only the two smallest apertures, 20$\arcsec$ and 40$\arcsec$,  are used for this survey since these observations are not sensitive to large-scale, smoothly varying extended emission (see \S \ref{2sig} and \S \ref{models}).  We used Starlink's ``aperadd'' package to measure the flux densities in instrument units in these apertures for both the sources and the calibrators.

In addition to the aperture FCFs, C$_{20}$ and C$_{40}$, we also calculate C$_{beam}$, the FCF for one beam (necessary for expressing both the peak flux of each source and the 1$\sigma$ RMS of the maps in units of Jy beam$^{-1}$).  Similar to the aperture FCFs, we calculated C$_{beam}$ by dividing the flux density of a calibrator in one beam by the value of the peak pixel of the map in the instrument units.  The flux density of the calibrator in one beam, S$_{beam}$, was calculated by assuming a gaussian beam.

Over the various observing runs, we used Mars, Uranus, and Neptune as calibrators.  As described above, we also used several secondary calibrators to check the pointing when none of the planets were available.  While they are adequate for checking the pointing, we did not use them as calibrators both because their flux densities are not as well known as for the planets and because they are not as bright as the planets and thus produce larger measurement uncertainties.  Table \ref{caltable} presents C$_{beam}$, C$_{20}$, C$_{40}$ for each observation of one of the three planets listed above.  Table \ref{caltable2} lists the average and standard deviation of the FCFs for each run.  It is clear from these tables that the FCFs are consistent both within and between each run.  We were unable to observe any of these three planets in March 2005, but since the FCFs are seen to be consistent between the various observing runs, we used the average FCFs over all of the runs (listed in Table \ref{caltable2}) to calibrate data taken during this run.  Based on the variation in FCF values from one observation to the next, we assign a 15\% calibration uncertainty to all of our data.  We investigated whether or not the beam sidelobe structure results in any dependence on the FCF values with the size of the calibrator, as the calibrators used here varied in size from $\sim 2.3$\as\ (Neptune) to greater than 10\as\ (Mars).  No significant variation is seen, and if such a variation is present at all, it is dominated by the 15\% calibration uncertainty.

\section{Results}\label{results}

The SHARC-II maps are oversampled, with pixel sizes three times smaller than those of a Nyquist sampled map.  We have used these oversampled maps to derive the properties listed in Table \ref{properties1}:  the Barycenter position\footnote{The Barycenter position of a source is defined to be the intensity-weighted center of the source.}, flux densities in 20\arcsec\ and 40\arcsec\ diameter apertures (S$_{20}$ and S$_{40}$, respectively), peak position, distance between Barycenter and peak positions ($\delta_{pk}$), and peak flux for each submillimeter source detected by SHARC-II.  For cores with no sources detected, we list the 3$\sigma$ upper limit in each aperture.  The Barycenter position is derived by extracting sources using Starlink's Extractor, which is based on SExtractor (Bertin 2003).  Extractor locates all sources above a specified intensity level, fits an ellipse to each source found, and calculates both the area ($A=\pi ab$, where a and b are the semi-major and semi-minor axes, respectively) and the aspect ratio of the ellipse ($\epsilon = a/b$).  The Barycenter position is derived by extracting sources at the 2$\sigma$ contours.  The flux densities and peak flux are derived as described in \S \ref{cal}, and the peak position is located by determining the position of the peak pixel.  The flux density uncertainties include components from both the measurement uncertainty and a 15\% calibration uncertainty, added in quadrature.  The uncertainty in peak flux also includes a 15\% calibration uncertainty.  The last column of Table \ref{properties1} lists whether or not each submillimeter source is associated with a \emph{Spitzer} source (see \S \ref{embedded}).  The peak fluxes (as well as the RMS noise of each map listed in Table \ref{info}) are given in units of Jy beam$^{-1}$.  To obtain these quantities in MJy sr$^{-1}$, a unit that can be used to compare between different instruments with different beams, the relevant conversions are:  1 Jy beam$^{-1}$ = 519.7 MJy sr$^{-1}$ at 350 $\mu$m; and 1 Jy beam$^{-1}$ = 310.3 MJy sr$^{-1}$ at 450 $\mu$m.

As stated in \S \ref{subobs}, the DSOS corrected all of the observations obtained in June 2005 for a zenith angle of $49\mbox{$^{\circ}$}$, regardless of the actual zenith angle of the observations.  All of these observations were obtained with actual zenith angles in the approximate range of $30-50\mbox{$^{\circ}$}$.  According to information presented on the DSOS website (see footnote to \S \ref{subobs} for web address), the corrections to the telescope efficiency introduced by the DSOS vary by up to approximately 10\% across this range of zenith angles.  Thus, 10\% is a conservative estimate of the additional uncertainty introduced by this equipment malfunction.  The flux density uncertainties for the June 2005 observations listed in Table \ref{properties1} include this additional 10\% uncertainty, added in quadrature to both the measurement uncertainty and 15\% calibration uncertainty.  The uncertainty in the peak flux for the sources observed in June 2005 also includes this additional 10\%.

For properties relating to the sizes and shapes of the cores, the oversampled maps may not always give reliable results.  Thus, we used the IDL procedure ``hrebin,'' available as part of the online IDL Astronomy User's Library\footnote{http://idlastro.gsfc.nasa.gov/}, to rebin the data into Nyquist sampled maps.  Sources were then extracted with Extractor, as described above, at both the half-maximum and 2$\sigma$ contours.  Table \ref{properties2} lists the results of these source extractions:  the major and minor axes, aspect ratio, and position angle of each submillimeter source, calculated at both the the half-maximum and 2$\sigma$ contours.

\subsection{2$\sigma$ Sizes}\label{2sig}

\begin{figure}[hbt!]
\epsscale{0.90}
\plotone{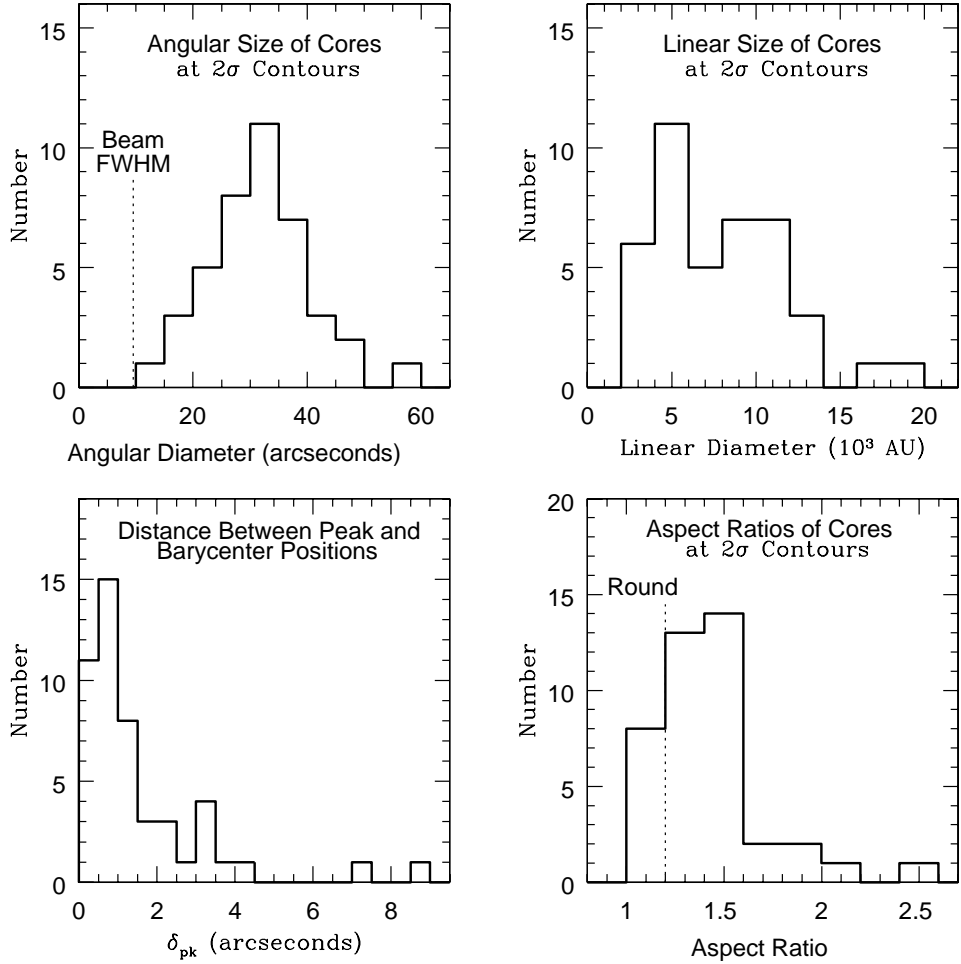}
\vspace{-1.0cm}
\caption{\label{hist}Histograms of core angular sizes at the 2$\sigma$ contours (upper left), linear sizes at the 2$\sigma$ contours (upper right), distances between Barycenter and peak positions (lower left), and aspect ratios at the 2$\sigma$ contours (lower right).}
\end{figure}

Figure \ref{hist} presents the distributions of the angular diameter at the 2$\sigma$ contour (defined to be $2\sqrt{ab}$, where $a$ and $b$ are the semi-major and semi-minor axes at the 2$\sigma$ contour, respectively), linear diameter at the 2$\sigma$ contour (defined simply as the angular diameter at the 2$\sigma$ contour multiplied by the distance to the core), distance between Barycenter and peak positions ($\delta_{pk}$), and aspect ratio ($\epsilon = a/b$) of each source.  As seen in Figure \ref{hist}, the distribution of 2$\sigma$ linear diameters has a mean and median of 8.1 $\times 10^3$ and 7.2 $\times 10^3$ AU, respectively, while the distribution of 2$\sigma$ angular diameters has both a mean and median of 32\as\ and a maximum of 57\as.  The 2$\sigma$ diameters derived from SHARC-II are generally $1.5-3$ times smaller than those derived from 850 \um\ SCUBA data (Young et al. 2006).  This discrepancy does not seem to correlate with the SHARC-II 2$\sigma$ angular diameter, it is observed in essentially all of the cores contained in both surveys regardless of the value of this quantity, including those with SHARC-II 2$\sigma$ angular diameters $\leq 20$\as.  Young et al. present their diameters at the 3$\sigma$ contours instead of the 2$\sigma$ contours, but since the sizes derived from these observations will be larger at the 2$\sigma$ contours than the 3$\sigma$ contours, this difference cannot explain the discrepancy.

\begin{figure}[hbt!]
\epsscale{0.90}
\plotone{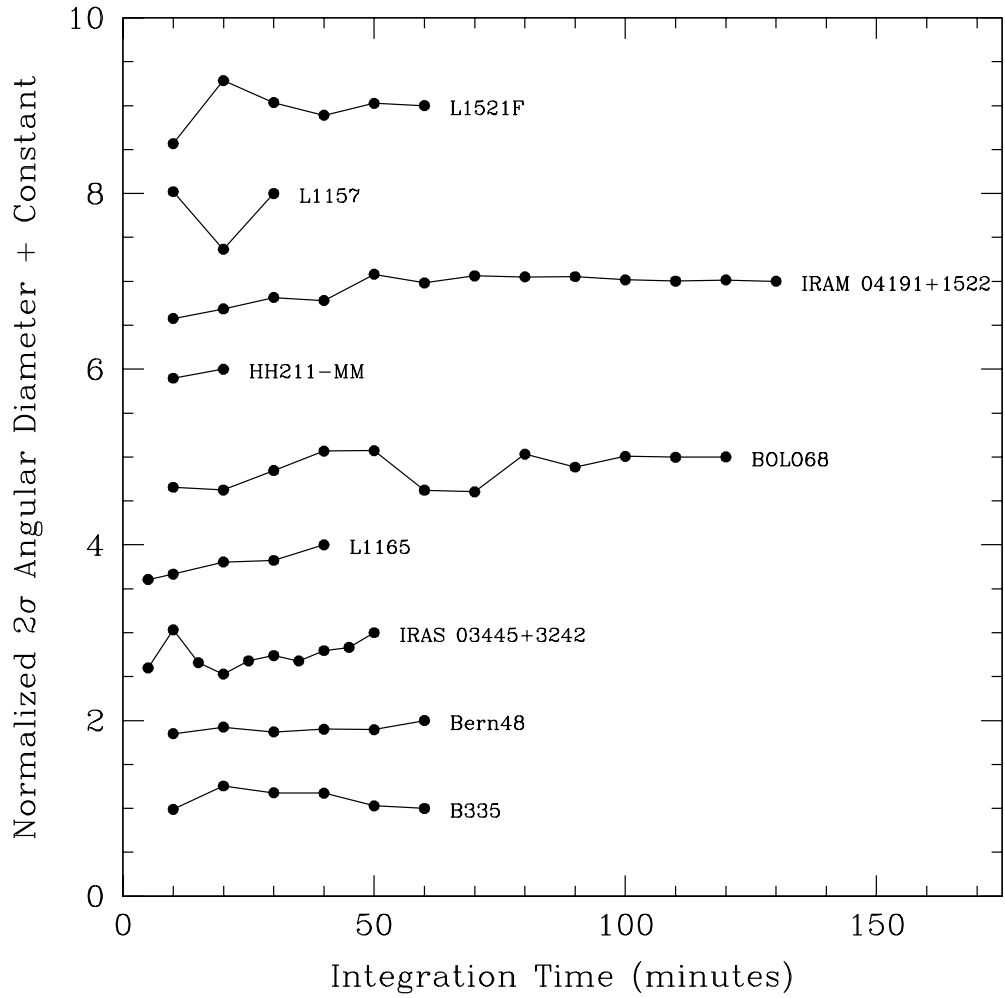}
\vspace{-1.0cm}
\caption{\label{sizesfig}2$\sigma$ angular diameters as a function of integration time for nine randomly selected 350 \um\ sources.  The data for each source has been normalized to the diameter at the full integration time, and offset by a constant to display on one plot.}
\end{figure}

One possible explanation for the difference in core sizes between these two surveys is that our observations may not go deep enough to detect the full extent of the core.  If so, the source size would increase with integration time, since more of the extended emission would be detected as the signal-to-noise ratio increased.  As stated in \S \ref{subobs}, the integrations on each core were separated into blocks of either five or ten minutes.  Thus, to test this possibility, Figure \ref{sizesfig} shows the angular diameter at the 2$\sigma$ contours as a function of integration time for a randomly selected subset of the 350 \um\ sources.

Some of the sources show slight rises in their 2$\sigma$ angular diameters as the integration time increases, but the trend does not continue past approximately 50 minutes of integration (IRAM 04191+1522 and BOLO68 being the best examples).  To consider a few specific examples, Young et al. (2006) derive a 3$\sigma$ diameter at 850 \um\ of 59\as\ for IRAM 04191+1522, while we derive a 2$\sigma$ diameter at 350 \um\ of 28\as.  As seen in Figure \ref{sizesfig}, integrating beyond about 50 minutes does not increase the size of this source.  As another example, Young et al. derive a 3$\sigma$ diameter at 850 \um\ of 111\as\ for L1521F, while we derive a 2$\sigma$ diameter at 350 \um\ of 34\as.  The 350 \um\ 2$\sigma$ diameter increases between 10 and 20 minutes of integration, but beyond that it does not increase with integration time.  Furthermore, many of the sources that show little to no change in their 2$\sigma$ angular diameters as the integration time increases do in fact have decreasing 1$\sigma$ RMS values as integration time increases (B335, Bern 48, and L1521F being good examples).  Thus, in general, we conclude that the discrepancy in source sizes cannot simply be attributed to the SHARC-II observations not going deep enough to detect the more extended emission seen in the SCUBA data.  An alternative explanation is thus required.

Two other possibilities may explain this discrepancy.  The first possibility is the fact that observations of low-mass cores at 350 \um\ are much more sensitive to the combination of temperature and density than those at longer submillimeter/millimeter wavelengths, which primarily only trace the dust column density.  In the Rayleigh-Jeans limit, the emission depends linearly on the dust temperature, but once this limit is violated the emission decreases exponentially with decreasing dust temperature.  At 850 \um, the Rayleigh-Jeans limit is satisfied for $T >> 17$ K, but at 350 \um, this limit is only satisfied for $T >> 41$ K.  Shirley et al. (2002) found a characteristic dust temperature for low-mass cores of $13.8 \pm 2.4$ K based on radiative transfer models of low-mass Class 0 protostars.  Thus, observations at 850 \um\ are marginally within the Rayleigh-Jeans limit, but observations at 350 \um\ are not.  The dust temperature of low-mass protostellar cores decreases from $\sim$ $50-300$ K in the inner regions of the cores\footnote{The range of temperatures arises from the range of protostellar luminosities, from $\leq 0.1$ \lsun\ for VeLLOs to $\sim$ $1-10$ \lsun\ for more luminous low-mass protostars} to $\sim$ $7-10$ K in the intermediate regions far away from the central source but not directly exposed to the Interstellar Radiation Field (e.g., Evans et al. 2001; Shirley et al. 2002; Young et al. 2003).  The 350 \um\ emission from these cores should thus decrease exponentially with increasing radius, meaning that the emission at this wavelength might decrease below the detection threshold much more quickly than 850 \um\ emission from the same core, resulting in smaller sizes derived from 350 \um\ data than from 850 \um\ data.

The second possibility is that the method used to obtain these SHARC-II observations might not be sensitive to extended emission.  Considering that both the mean and median of the distribution of 2$\sigma$ angular diameters is 32\as\ and that the SHARC-II field of view is 2\farcm59 $\times$ 0\farcm97, the observations presented in this work appear to lose sensitivity to extended emission on scales larger than about half the size of the smaller dimension of the array.  This is in qualitative agreement with the statement from \S \ref{subobs} that the Lissajous observing method used for this survey works best for sources with sizes less than or comparable to the size of the array.  It also agrees with the findings of Enoch et al. (2006), who showed that their observations of low-mass cores in Perseus with the Bolocam bolometer array at the CSO could not recover extended emission on size scales larger than about half the size of the array, and while the details of the data collection and reduction between Bolocam and SHARC-II are different, they are qualitatively similar.  A lack of sensitivity to extended emission is not surprising since, as stated in \S \ref{subreduction}, sky emission is subtracted to first order by removing emission common to all bolometers.  Thus, extended emission larger than the size of the array cannot be completely separated from sky emission.  Finally, such a discrepancy in source sizes between observations with different instruments is not unique to this survey; low-mass cores observed at both 850 \um\ with SCUBA and 1.2 mm with MAMBO show significantly (between factors of $2-20$) larger 2$\sigma$ angular diameters at 1.2 mm than 850 \um\ (Young et al. 2006; J. Kauffmann et al. 2006, in preparation).  Since both 850 \um\ and 1.2 mm are within the Rayleigh-Jeans limit, temperature effects cannot account for these discrepancies.

\begin{figure}[hbt!]
\epsscale{0.90}
\plotone{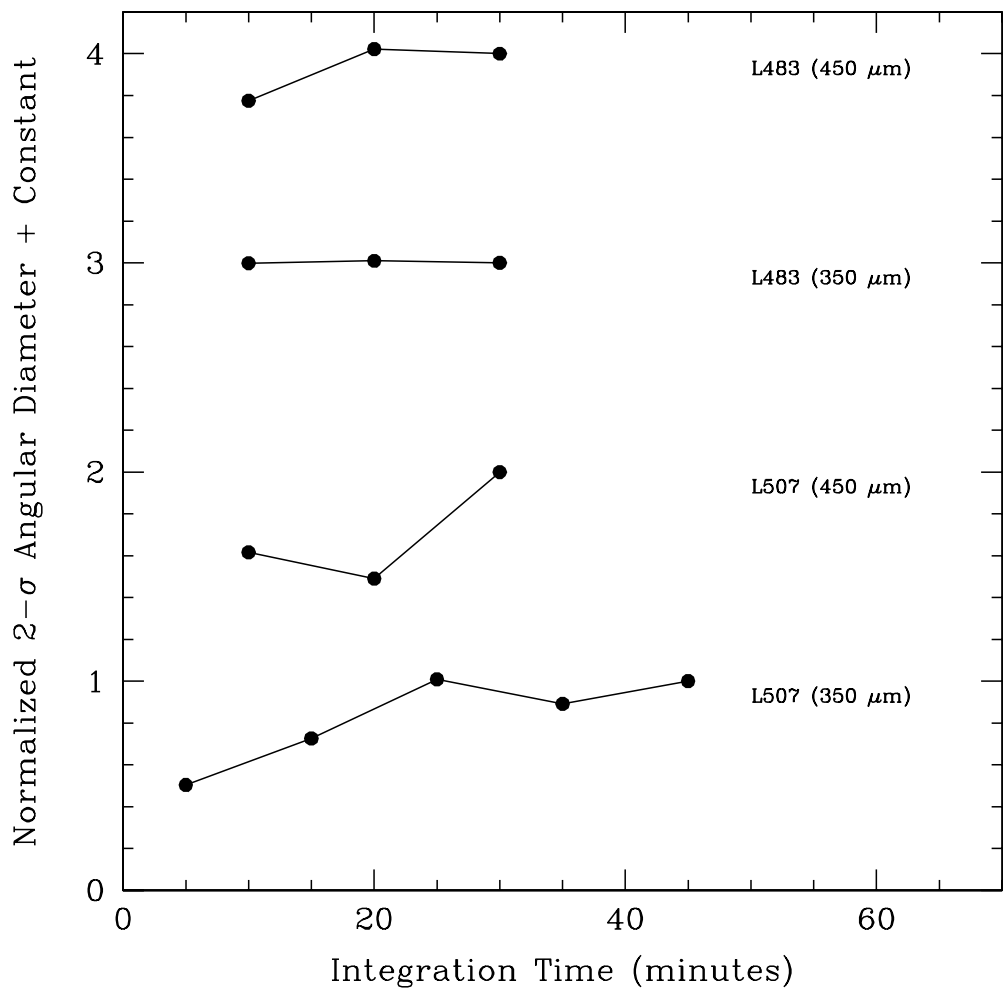}
\vspace{-1.0cm}
\caption{\label{sizesfig2}Same as Figure \ref{sizesfig}, except for the two cores observed at both 350 and 450 \um.  The fact that the 450 \um\ 2$\sigma$ angular diameter for L507 increases when the last scan is added suggests that the integration on this source is not yet deep enough to use in comparison with the 350 \um\ 2$\sigma$ angular diameter.  The opposite is true for L483, the integrations on this source are deep enough at both 350 and 450 \um\ to compare the derived 2$\sigma$ angular diameters at these two wavelengths.}
\end{figure}

In reality, the discrepancy between source sizes is most likely a combination of the temperature and instrumental effects discussed above.  Comparing the derived 2$\sigma$ angular diameters for sources observed with the same instrument at 350 and 450 \um\ could help distinguish between these two possibilities.  However, only two cores were observed at both wavelengths, and, as Figure \ref{sizesfig2} shows, only L483 has deep enough integrations at both wavelengths to be used for this purpose.  We measure a 2$\sigma$ angular diameter for L483 of 47\as\ at both 350 and 450 \um.  A much larger sample is required to draw any conclusions, but the exact agreement between these values, combined with the above discussion, leads us to conclude that instrumental effects are primarily responsible for the discrepancy in source sizes.  However, the fact that this discrepancy is observed in even the most compact SHARC-II sources suggests a more subtle effect than simple insensitivity to extended emission beyond a fixed size scale.  We will return to this question of sensitivity to extended emission in \S \ref{models}, but we note here the basic conclusion from that section that the observations presented in this survey are insensitive to smoothly varying extended emission.

An inability to completely recover extended emission has important implications for the flux densities in 40\as\ apertures presented in Table \ref{properties1}.  Furthermore, the fact that several sources in Figures \ref{sizesfig} and \ref{sizesfig2} with less than 50 minutes of total integration time show increasing diameters with integration time right up until the full integration suggests that longer integration times may be required on such sources to reach the size scale at which the SHARC-II array loses sensitivity.  Both the flux densities in 40\as\ apertures presented in Table \ref{properties1} and the major and minor axes at the 2$\sigma$ contours presented in Table \ref{properties2} should be used with the important caveat that they most likely do not capture the full emission and extent of the cores.

\subsection{FWHM}\label{fwhm}

\begin{figure}[hbt!]
\epsscale{0.90}
\plotone{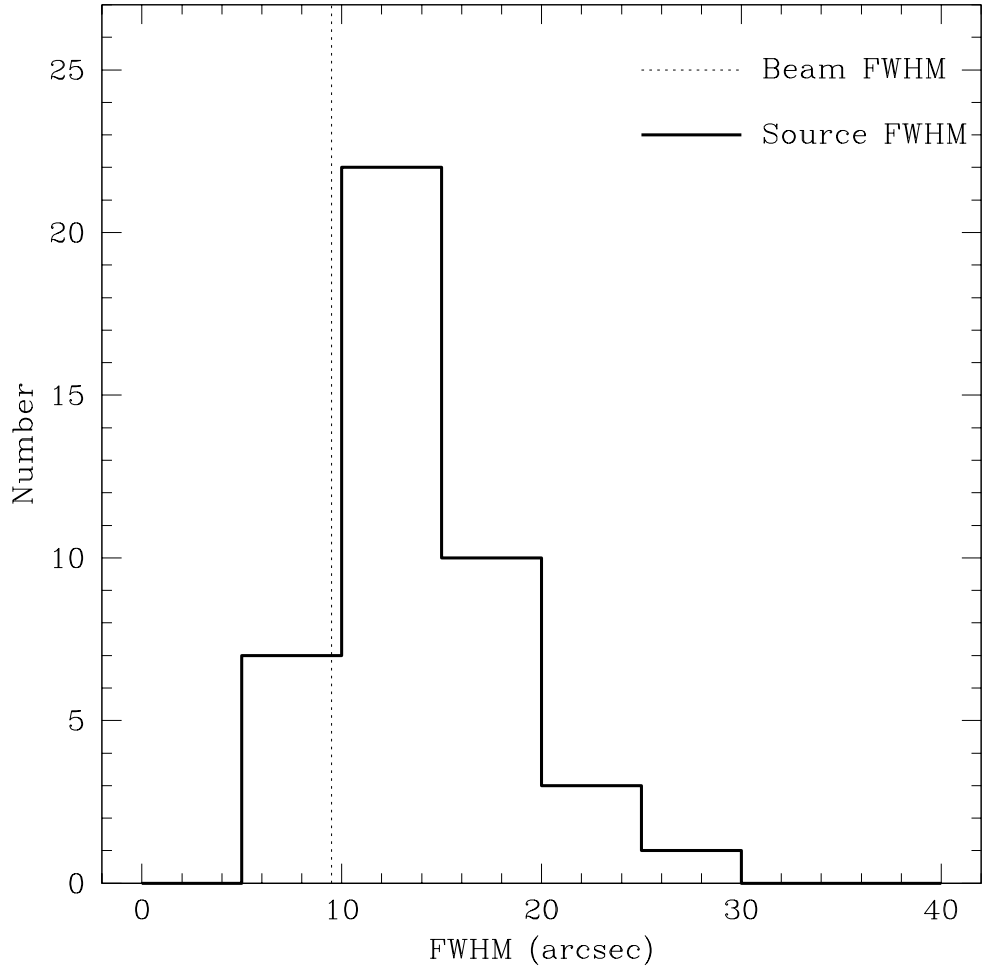}
\vspace{-1.0cm}
\caption{\label{fwhmfig}Distribution of FWHM sizes of 350 \um\ sources.}
\end{figure}

Figure \ref{fwhmfig} shows the distribution of the Full-Width Half-Maximum (FWHM) for the 350 \um\ sources.  The distribution has a mean and median of 13\farcs3 and 12\farcs3, respectively, and shows a narrow distribution similar to that seen in M. Enoch et al. (2006, in preparation) for sources in Serpens observed at 1.1 mm with Bolocam.  Using models simulating the collapse of a singular isothermal sphere (the ``standard model''; Shu, Adams, \& Lizano 1987), Terebey et al. (1993) showed that the FWHM of a source depends on the size of the beam.  More recently, Young et al. (2003) showed that, for observations of cores with power-law radial density profiles, the ratio of the deconvolved source size\footnote{The deconvolved source size, $\theta_{dec}$, is determined from the FWHM size of the beam ($\theta_{mb}$) and source ($\theta_{src}$) intensity profiles:  $\theta_{dec}=(\theta^2_{src}-\theta^2_{mb})^{1/2}$} to the beam size is correlated with the index of the density power law.  Assuming a beam FWHM of $\theta_{mb}=$8\farcs5, the mean source FWHM of $\langle$$\theta_{src}$$\rangle=$ 13\farcs3 corresponds to a mean deconvolved source size of $\langle$$\theta_{dec}$$\rangle=$ 10\farcs2.  This gives a mean ratio of $\langle$$\theta_{dec}/\theta_{mb}$$\rangle=$ 1.2.  Placing this value on Figure 27 of Young et al. (2003), which plots the index of the density power-law, $p$, versus the ratio of the deconvolved source size to beam size, $\theta_{dec}/\theta_{mb}$, we conclude that the average value of the density power-law index is $\langle$$p$$\rangle$ $\sim$ 1.8.

This calculation is only meant to serve as a quick approximation to the average value of $p$ for the cores observed in this survey; full radiative transfer models including data at longer wavelengths would be required to truly determine $p$, a task beyond the scope of this paper.  However, we note that a value of $\langle$$p$$\rangle$ $=$ 1.8 is identical to the median value of $p$ determined by Young et al. (2003) through radiative transfer models of nine Class I protostars.  Furthermore, the standard model predicts a collapsing envelope with $p=1.5$ surrounded by a static region with $p=2.0$, with the transition between the two propagating outward at the sound speed.  Thus, an average power-law index of $\langle$$p$$\rangle$ $\sim$ 1.8 for the cores in this survey is not a surprising result.  Since the cores in this survey are located at different distances, this result suggests the power law in density extends over a large linear size range (a mean deconvolved source size of 10\farcs2 over the distance range of $125-450$ pc for the cores in this sample corresponds to a linear size range of $\sim$ $1000-4500$ AU), as predicted by many models of star formation, including the standard model.

\subsection{Shapes}\label{shapes}

\begin{figure}[hbt!]
\epsscale{0.90}
\plotone{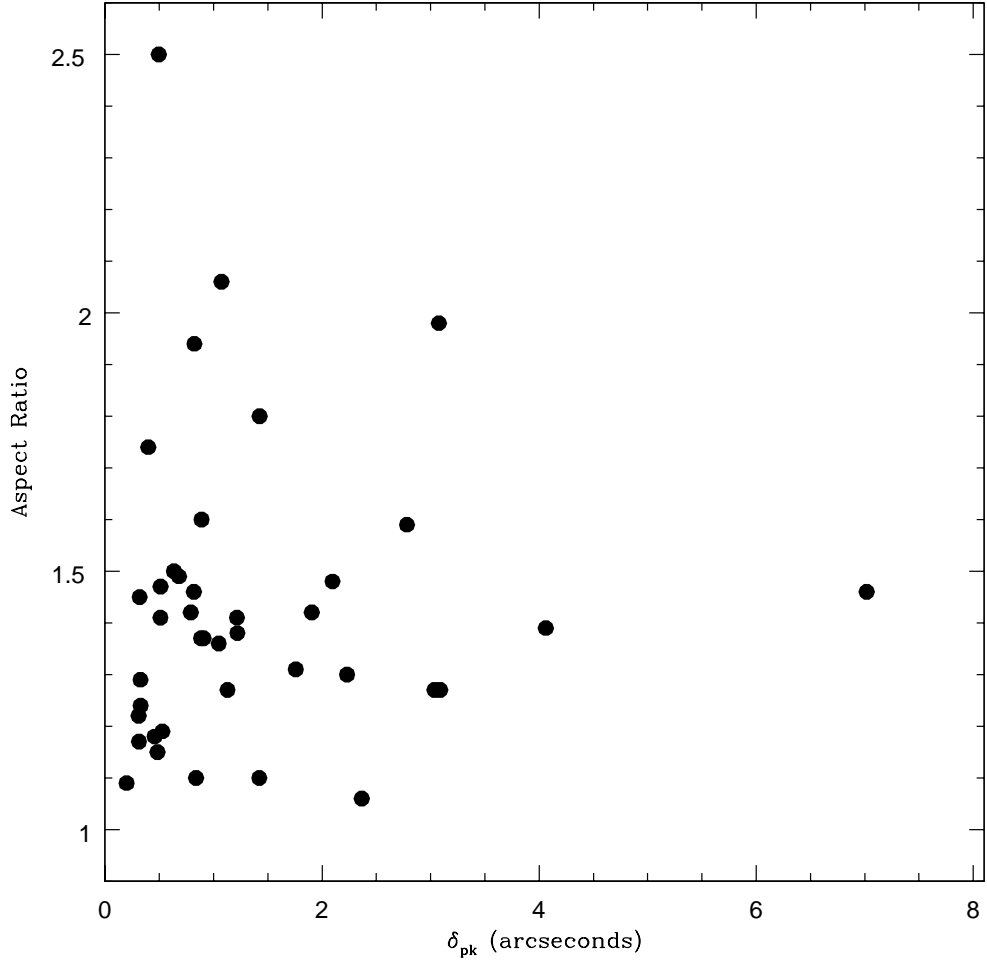}
\vspace{-1.0cm}
\caption{\label{deltapk}Aspect ratio of each core plotted as a function of $\delta_{pk}$, the distance between the Barycenter and peak positions of the core.}
\end{figure}

The distance between the peak and Barycenter positions, $\delta_{pk}$, is a measure of the degree of axial symmetry of each source, while the aspect ratio is a measure of the roundness of the source.  Following Enoch et al. (2006), we consider any source with an aspect ratio less than 1.2 to be round.  The distribution of aspect ratios in Figure \ref{hist} shows that, while most sources do have aspect ratios at the 2$\sigma$ contours greater than 1.2, very few show large flattening (only two sources have aspect ratios greater than 2.0).  Figure \ref{deltapk}, which plots the aspect ratio of each source as a function of its value of $\delta_{pk}$, shows no significant correlation between roundness and axial symmetry.

\subsection{Multiplicity}\label{mult}

Nine of the 41 cores detected by SHARC-II (approximately 20\%) contain multiple submillimeter sources:  L1455, IRAM 04191+1522, B18-4, B35A, CG30, L43, L328, L1221, and L1251B.  The multiple sources in L1455, IRAM 04191+1522, B18-4, L43, L1221, and L1251B all correspond to previously known sources, although the eastern, starless core in L43 is resolved into several submillimeter peaks that were not previously known.  The multiplicity in B35A, CG30, and L328, however, is a new result.  As 350 $\mu$m observations are more sensitive to temperature than longer-wavelength bolometer observations, multiple protostellar sources contained in one core could result in localized heating of the dust surrounding each protostar and thus the detection of multiple sources at 350 $\mu$m, but the detection of only one source at the longer wavelengths that are more sensitive to density than temperature.  Indeed this does seem to be the case for CG30, as both of the detected 350 \um\ sources are associated with embedded Young Stellar Objects (see \S \ref{embedded}).  However, one of the three sources in B35A and two of the three sources in L328 are not associated with embedded sources (\S \ref{embedded}), and, furthermore, the starless core in L43 is resolved into multiple sources despite having no embedded sources.  The exact sensitivity of \emph{Spitzer} to embedded, Very Low Luminosity Objects (VeLLOs) is still under investigation, but objects with luminosities $\leq 0.1$ \lsun\ are easily detected (e.g. Young et al. 2004; M. Dunham et al. 2007, in preparation), ruling out the presence of an undetected VeLLO to a high degree of confidence.  Thus, the new multiplicity seen in these maps cannot be explained solely by heating from an embedded source.

Instead, the lack of embedded sources in some of these multiple sources suggests that, for these sources, multiple density peaks are in fact present.  Since the beam size of 8\farcs5 for SHARC-II observations at 350 \um\ is significantly smaller than typical beam sizes for longer-wavelength surveys (e.g., 15\farcs5 for SCUBA observations at 850 \um; Young et al. 2006), the SHARC-II observations provide higher spatial resolution.  Thus, it is not surprising that these observations may resolve some cores into multiple 350 \um\ sources even when no embedded sources are present.  As discussed in \S \ref{starless}, the observations presented in this work do not appear very sensitive to cores without embedded sources, but all of the new 350 \um\ sources without embedded protostars are located close to other protostars and thus other sources of heating, and are thus likely to be warmer than typical isolated starless cores.

Ultimately, sorting out temperature effects from density effects when investigating the multiplicity presented here will require comparing these results to data of comparable resolution but less sensitivity to temperature.  Either much higher-resolution data at longer wavelengths than currently available, or, ideally, high-resolution extinction maps are needed to fully investigate the physical structure of these three cores with newly detected multiple sources that aren't all associated with embedded protostars.

\subsection{Association with Embedded Young Stellar Objects}\label{embedded}

The last column of Table \ref{properties1} indicates whether or not each submillimeter source is associated with an embedded Young Stellar Object (YSO) based on a search of \emph{Spitzer} c2d data.  Based on the c2d data delivered to the Spitzer Science Center in December 2005 (Evans et al. 2005), the c2d team identified candidate YSOs based on their positions in two different \emph{Spitzer} color-magnitude diagrams:  $[8.0]$ vs. $[4.5]-[8.0]$ and $[24]$ vs. $[8.0]-[24]$.  The details of this identification of candidate YSOs can be found in Harvey et al. 2006 and J\o rgensen et al. 2006, but to summarize, a source had to meet at least one of the two following sets of criteria to be classified as a candidate YSO in the 2005 catalogs:
\begin{itemize}
\item $[4.6]-[8.0] > 0.5$ and $[8.0] < 14-([4.6]-[8.0])$
\item $[8]-[24] > 0.7$ and $[24] < 12-([8]-[24])$
\end{itemize}
These criteria were selected both to pick out red objects with excess infrared emission over that of stars and to eliminate most galaxies, which can appear similar to Young Stellar Objects in \emph{Spitzer} data.  It is important to note that these criteria were selected to eliminate galaxies in a statistical sense and do not completely separate YSOs from galaxies.  The c2d team is developing a revised set of criteria that allow for a more complete separation (P. Harvey et al. 2006, in preparation; Porras et al. 2006), but the original criteria are sufficient for this study since we also visually inspect each source (see below).  We searched for objects classified as candidate YSOs, based on the above criteria, within 10\as\ (approximately 1 beam) of the peak position of each submillimeter source.  For cores with no detections, we searched for candidate YSOs within the full region covered by the SHARC-II map since peak positions are unavailable for these observations.  We also visually inspected each source to verify whether or not it contained a Young Stellar Object since these criteria do not completely separate YSOs from galaxies.  We note in the table whether or not a YSO was found for each submillimeter source, and we also indicate the cores for which \emph{Spitzer} data are unavailable.  There were four cases where the presence or absence of an embedded YSO as indicated in Table \ref{properties1} was changed after visual inspection.  We note these cases in the table and describe them below.

The first case is L1455-IRS2.  There is no candidate YSO detected by \emph{Spitzer} within 10\as\ of the peak position of L1455-IRS2, but there is one 11\farcs9 away.  The 350 $\mu$m map suggests L1455-IRS2 may in fact be comprised of two separate submillimeter sources, and although they are not well-resolved enough to present them as two separate sources, the \emph{Spitzer} source is coincident with the western sub-structure.  Furthermore, L1455-IRS2 is the weakest of the four sources in the 350 \um\ map of L1455, and considering that J\o rgensen et al. (2006) concluded that L1455-IRS2 has mid-infrared colors consistent with those of a Class II object, it is not surprising that it is associated with weak 350 \um\ emission. Based on these data, we consider L1455-IRS2 to be associated with an embedded YSO despite not meeting the above criterion of being located less than 10\as\ from the peak of the emission.  An alternative possibility is that the \emph{Spitzer} source and the 350 \um\ emission are seen together in projection but not actually associated with each other.  We consider this to be less likely but are unable to rule it out.

Another case is HH211-MM.  HH211 was originally discovered in near-infrared H$_2$ emission by McCaughrean et al. (1994), and HH211-MM is a millimeter source known to be the driving source of a corresponding CO outflow (Gueth \& Guilloteau 1999) that has recently been studied in multiple transitions of SiO and CO at high angular resolution with the SMA (Hirano et al. 2006; Palau et al. 2006).  Rebull et al. (2006) show that there is a corresponding \emph{Spitzer} source detected only at 70 \um, placing it among the reddest objects in the Perseus cloud.  It is not classified as a candidate YSO since detections at \emph{Spitzer} wavelengths shortward of 70 \um\ are required for such a classification, but the detection at 70 \um\ and the fact that this source is driving an outflow lead us to conclude that HH211-MM is associated with an embedded source.

L43-SMM3 is the third case.  There is an object detected by \emph{Spitzer}, located $\sim$ 9\as\ from the peak position of L43-SMM3, that has 4.5 and 8.0 $\mu$m fluxes consistent with being a candidate YSO.  However, it is not detected by \emph{Spitzer} at 24 or 70 $\mu$m.  The combination of the fact that the sample of candidate YSOs identified with \emph{Spitzer} data are expected to contain a small amount of contamination from galaxies, as described above, and the nondetections at 24 and 70 $\mu$m, lead us to conclude that this object is unlikely to actually be a protostar.  Thus, we do not consider L43-SMM3 to be associated with an embedded YSO.

The final case is L1251A.  There are \emph{Spitzer} sources within 10\as\ of both L1251A-1 and L1251A-2 that fail to meet the criteria for classification as a candidate YSO but still appear to be protostars.  The source associated with L1251A-1 was not detected at 8.0 $\mu$m, and since a detection at this wavelength is required according to the above criteria, it is not classified as a candidate YSO.  The source associated with L1251A-2 falls just beyond the $[24] < 12-([8]-[24])$ criterion, but since these criteria provide a statistical sample of candidate YSOs that do not separate protostars and galaxies with complete reliability, the failure to meet this criterion does not rule out the possibility that this source is in fact a YSO.  Additionally, both sources are detected by \emph{Spitzer} at 70 $\mu$m; thus we consider L1251A-1 and L1251A-2 to be associated with embedded YSOs.

\section{Discussion}\label{starless}

\subsection{Detection vs. Nondetection}\label{detnondet}

Of the 53 cores observed and listed in Table \ref{info}, 12 are classified as starless based upon no detection of an embedded protostar by either \emph{IRAS} or \emph{Spitzer}:  L1521B, L1521E, B18-1, TMC2, B18-4, TMC1-A, L1544, L134A, L43\footnote{L43 contains both a starless core and a protostar}, L492, L694-2, and L1021.  Only three of these 12 starless cores are detected by SHARC-II:  B18-4, L43, and L492.  Both B18-4 and L43 are located close to potential strong sources of external heating, IRAS 04325+2402 in the case of B18-4 and L43-RNO91 in the case of L43.  Furthermore, the three starless sources in L43 are located near the edge of the blue-shifted outflow emission from L43-RNO91, as seen from the CO (2-1) maps presented in Bence et al. (1998).  In fact, the sharp western edge of these three sources, extending from the southeast to the northwest, directly correlates with the edge of this blue-shifted outflow emission.  A similar situation exists for B18-4; IRAS 04325+2402 shows extended, bipolar emission in near-infrared HST images that suggests an outflow pointing in the direction of B18-4 (Hartmann et al. 1999).  A different direction for the IRAS 04325+2402 outflow is suggested from CO data (extending to the northwest; Heyer et al. 1987), but it is not clear this CO outflow is actually associated IRAS 04325+2402 (see Hartmann et al. 1999 for discussion).  Since an outflow will heat any dust condensation it encounters, the only two starless cores detected in this survey are most likely warmer than would be expected for an isolated, starless core.

The same can not be said for L492, but the detection of L492 is quite weak ($\sim 5 \sigma$).  None of the three detected starless cores show very centrally peaked emission.  Of the nine starless cores not detected, all have $1\sigma$ noise values well below the average over all the maps presented in this sample, suggesting that these non-detections cannot simply be attributed to lower sensitivity than the rest of the sample.  However, since the cores observed in this survey are located at different distances, a more careful analysis is required before such a claim can be made.

To account for possible distance effects, we have converted the measurement of the 1$\sigma$ RMS for each core into distance-independent units of \msun\ beam$^{-1}$.  The sensitivity of a given map in these units, for a core with a 1$\sigma$ RMS of $\sigma$, at distance $d$ and an isothermal dust temperature $T$, is given by the relation
\begin{equation}
\sigma_{M}=\frac{d^2 \sigma}{B_{\nu}(T_d) \kappa_{\nu}},
\end{equation}
where $\kappa_{\nu}=0.101$ cm$^2$ gm$^{-1}$ is the dust opacity at 350 \um\ from Ossenkopf \& Henning (1994) divided by the assumed gas-to-dust mass ratio of 100.  The last column of Table \ref{info} presents this quantity for all the cores in this survey, assuming an isothermal dust temperature of $T=15$K.  In general, the isothermal approximation is not valid for deriving masses from these observations since the 350 \um\ emission from low-mass cores, and through this the calculated masses, will depend exponentially on the assumed temperature (see discussion in \S \ref{2sig}).  It is for this reason that we do not derive the masses of the cores in this survey; that is a task best left to surveys at longer wavelengths.  However, since all we are interested in here is the relative sensitivity between maps, this approximation is valid; changing the value of $T_d$ will change all of the values of $\sigma_{M}$ by the same factor.  This strong dependence on temperature is an important caveat to keep in mind, however, before using the tabulated $\sigma_{M}$ as absolute rather than relative sensitivities.

Based on the values for $\sigma_{M}$ given in Table \ref{info}, all 12 starless cores in this survey have sensitivities below the average for the entire survey.  This is not surprising since the observing strategy pursued was to integrate until we obtained either a strong detection or a strong upper limit, and it leads us to conclude that the non-detections of 9 out of the 12 starless cores observed cannot be attributed to lower sensitivities than the rest of the sample.

In addition to the nine starless cores not detected, the other three cores not detected by SHARC-II are LkH$\alpha$ 327, DC255.4-3.9, and L1148B.  Four other cores, IRAS 03301+3111, IRAS 03439+3233, EC74, and EC88, show very weak detections (between 3$\sigma$ and 6$\sigma$). Excluding DC255.4-3.9 and L1148B, which are discussed below, all of these cores were selected from IRS targets of interest that were not restricted to objects known to still be heavily embedded in dense cores.  The distance-independent sensitivities obtained for these objects are all at or slightly above average, but are well in line with the typical sensitivies obtained for the rest of the survey.  Thus, weak or non-existent submillimeter detections for these cores can be used to infer that they are in later stages of evolution where most of the dense, circumstellar envelope has dissipated.  Because of their likely status as more evolved objects, they are not included in the discussion below.

\subsection{Starless vs. Protostellar Cores}\label{starlessproto}

The detection statistics quoted above suggest that the SHARC-II observations presented in this paper distinguish quite well between protostellar cores and isolated, starless cores.  This is not unexpected because, as already mentioned, 350 $\mu$m is much more sensitive to the dust temperature than is emission at longer wavelengths.  Thus, the presence or absence of central heating from an embedded protostar should be more noticeable at this wavelength.  This is explored further with simple radiative transfer models in \S \ref{models}, but, qualitatively, starless cores are generally not detected while protostellar cores generally are detected.  Two of the three exceptions to this appear to occur when there is a strong souce of external heating, as in the case of B18-4 and L43, and all three of the exceptions show much less centrally peaked emission than cores with protostars.  This comes with the important caveat, however, that only 12 starless cores have been observed, and seven of them are located in Taurus.  L492, the only starless core detected that is not close to an obvious source of strong external heating, is also the only starless core observed that is located in the Serpens molecular cloud.  The preliminary results from this survey suggest that a clear distinction between starless cores and those with protostars is present in SHARC-II observations.

Perhaps the most interesting result from this survey concerns the VeLLOs.  As discussed in \S \ref{intro}, \emph{Spitzer} c2d observations have shown that several cores believed to be starless (containing no protostars) actually harbor Very Low Luminosity Objects (VeLLOs) with internal luminosities $L_{int} \leq 0.1$ \lsun\ (e.g. Young et al. 2004,  Di Francesco et al. 2006).  Three confirmed VeLLOs (IRAM 04191+1522, Dunham et al. 2006; L1521F, Bourke et al. 2006; L1014, Young et al. 2004) and seven candidate VeLLOs (DC255.4-3.9, CG30, L507, L328, L1148B, L1221, and L673-7) are included in the survey\footnote{Confirmed VeLLOs have had their very low internal luminosities confirmed with radiative transfer models, while candidate VeLLOs have simply been identified as potential Very Low Luminosity Objects based on their \emph{Spitzer} fluxes (T. Huard et al. 2006, in preparation).}.  Except for DC255.4-3.9 and L1148B, which are not detected, VeLLOs show reasonably strong detections and centrally peaked emission, reminiscent of cores with protostars except with somewhat lower integrated flux densities.  In other words, VeLLOs look much more like cores with protostars than starless cores in SHARC-II data, indicating that observations of VeLLO candidates with SHARC-II can be an essential part of confirming their status as very low luminosity, embedded objects.  The nondetections of DC255.4-3.9 and L1148B cast doubt on their status as embedded objects, since embedded objects appear to be essentially always detected.  Future study will be devoted to these and other VeLLOs (J. Kauffmann et al. 2006, in preparation; M. Dunham et al. 2007, in preparation).

\subsection{Simple Radiative Transfer Models}\label{models}

In an effort to better understand the conclusion that a clear distinction between starless and protostellar cores is seen in these observations, we have used the one-dimensional radiative transfer code Dusty (Ivezic et al. 1999) to construct simple models representative of each of the three types of cores observed in this survey: starless cores, cores harboring VeLLOs, and cores harboring more luminous protostars ($\sim 1$ \lsun).

For the model representative of the VeLLOs, we used that of L1521F described in detail in Bourke et al. (2006): a 4.8 \msun\ envelope heated both internally by a protostar with $L_{int} = 0.05$ \lsun\ and externally by the Interstellar Radiation Field (ISRF).  For the model representative of the starless cores, we simply modified the model for L1521F to include only external heating by the ISRF, and for the model representative of the cores with more luminous protostars, we modified the model for L1521F by increasing the internal luminosity of the central source from $L_{int} = 0.05$ \lsun\ to $L_{int} = 1.0$ \lsun.  All models are placed at an assumed distance of 140 pc, the distance to Taurus.  We then calculated the expected flux densities from these three models in 20\as\ and 40\as\ apertures.  In 20\as\ apertures, the models predicted flux densities of 1.23, 2.86, and 12.9 Jy for the starless, VeLLO, and 1 \lsun\ protostellar cores, respectively.  In 40\as\ apertures, the predicted flux densities are 4.48, 6.86, and 26.0 Jy.

Comparing the flux densities predicted by the model of a VeLLO core with the observed flux densities of L1521F shows that they are in good agreement with each other, as expected since the representative VeLLO model assumed here is exactly the L1521F model from Bourke et al. (2006) that provided the best fit to the observed SED, including data from this survey.  The envelopes of L1521F and L1544 are, generally speaking, quite similar to each other.  Using 1.2 mm continuum maps, Crapsi et al. (2005) defined the quantity $r_{70}=\sqrt{A_{70} / \pi}$, where $A_{70}$ is the area within the 70\% contour of the dust peak, and showed that both L1521F and L1544 have $r_{70} < 4800$ AU.  Thus, both cores have high degrees of central condensation, and, in fact, these two cores were selected by Crapsi et al. as being the most ``evolved'' starless cores in their sample (the protostar embedded within L1521F was not yet discovered).  Furthermore, detailed radiative transfer models show that similar envelope parameters provide good fits to the observations of both cores (e.g., Crapsi et al. 2004; Bourke et al. 2006; Evans et al. 2001; Doty et al. 2005).  Finally, both L1521F and L1544 are located in Taurus, and should be exposed to a similar amount of external heating from the ISRF.  Thus, the model assumed in this work as being representative of starless cores should be a fairly good approximation for L1544, and, in fact, this is the reason we selected this particular model.

Table \ref{l1544model} presents both the modeled and observed flux densities for L1544 from $90-1300$ \um, using the simple model described above and observations from both this survey and from the literature (Ward-Thompson et al. 2002; Shirley et al. 2000).  For all wavelengths excluding 350 \um, the agreement between the observations and this simple model, as well as the agreement between these model predictions and those of the more detailed models of L1544 by Evans et al. (2001), confirm that it is a good approximation for L1544.  The slight increase in modeled over observed flux densities at 850 and 1300 \um\ can be attributed to the fact that the L1521F envelope is more massive than L1544 by approximately a factor of 2 (Bourke et al. 2006; Doty et al. 2005).  At 350 \um, however, the model and observations from this survey disagree by nearly two orders of magnitude.

Differences of a factor of a few could be explained by the fact that the starless core model is not exactly a model of L1544, but, in order to match the model to the SHARC-II 350 \um\ observation of L1544, either the strength of the ISRF or the envelope parameter that sets the scale for the density would have to be decreased to unreasonably low values.  Since we can eliminate these possibilities based on the above confirmation that this model is a good approximation for L1544, we turn our attention to instrumental effects.

\begin{figure}[hbt!]
\epsscale{0.90}
\plotone{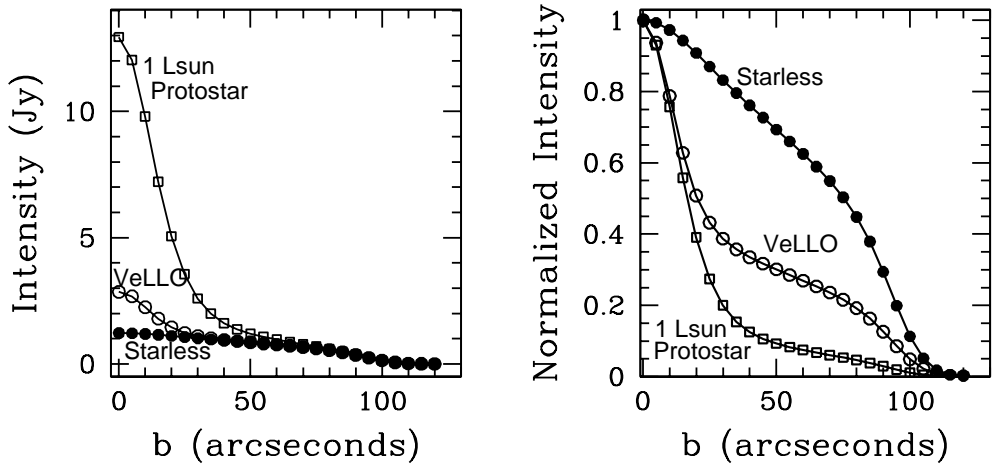}
\vspace{-1.0cm}
\caption{\label{profiles}Intensity profiles for models representative of the three types of cores observed in this survey: protostellar cores harboring $\sim 1$ \lsun\ protostars (solid line with open squares), protostellar cores harboring VeLLOs (solid line with open circles), and starless cores (solid line with filled circles).  The intensity profiles are displayed on both un-normalized (\emph{left}) and normalized (\emph{right}) scales.}
\end{figure}

Figure \ref{profiles} presents intensity profiles for all three representative models, on both un-normalized and normalized scales.  As expected, the core harboring the 1 \lsun\ protostar is the brightest and the starless core is the weakest.  However, displaying the intensity profiles on a normalized scale shows a striking contrast between protostellar and starless cores, regardless of the actual luminosity of the internal source.  While the absolute scale of the emission is different, the emission both from cores harboring 1 \lsun\ protostars and from cores harboring VeLLOs falls below 40\% of the maximum within the central 30\as, and in fact the two normalized intensity profiles are nearly indistinguishable within the central 20\as.  The emission from the starless core, on the other hand, decreases by only about 15\% from the maximum within 30\as, and less than 10\% within 20\as.  Furthermore, we note that the 350 \um\ emission from both types of protostellar cores drops off much less rapidly beyond $\sim 30-40$\as, exactly the scale at which the evidence presented in \S \ref{2sig} suggests our observations lose sensitivity to extended emission.

Ultimately, we conclude that the observations presented in this survey are insensitive to smoothly varying extended emission, as seen by both the inability to detect most starless cores and the small sizes dervived for the detected cores.  While a quantitative analysis of the size scales over which extended emission must vary in order to be separated from sky emission is beyond the scope of this paper, we note that it must be greater than 15\% over 30\as\ based on the non-detection of L1544.  Future work is needed to develop a better understanding of these instrumental effects.

\section{Conclusions}\label{conclusions}

We present maps of 53 low-mass dense cores with SHARC-II at the CSO, 52 at 350 $\mu$m and 3 at 450 $\mu$m, with two observed at both wavelengths.  41 of these cores are detected while 12 are not, and 9 of the 41 detected cores show multiple submillimeter sources.  We derive and tabulate the basic properties for each detected submillimeter source:  position, fluxes in 20\as\ and 40\as\ apertures, peak flux, size, aspect ratio, and position angle.  We also use data from the \emph{Spitzer Space Telescope} to indicate whether or not each core is associated with an embedded Young Stellar Object.

The sizes of the cores as derived from these observations are typically smaller than sizes of the same cores derived from other, longer-wavelength observations by about a factor of $1.5-3$.  This is not simply a case of not integrating long enough to detect the extended emission; instead it likely arises primarily from the fact that the observations presented in this survey are insensitive to smoothly varying extended emission.  Future investigation is needed to develop a better understanding of these instrumental effects, and care should be taken in using the data presented in this survey.  In particular, the upper limits presented for the starless cores not detected should be considered upper limits based on the type of emission these observations are capable of detecting and not true upper limits of the flux densities at 350 \um.  Some care should also be exercised in using the flux densities presented in 40\as\ apertures for the detected sources.

The results of this survey suggest that SHARC-II observations of dense cores are capable of distinguishing between starless cores and cores with protostars much better than observations with other bolometer arrays at longer wavelengths.  VeLLOs, Very Low luminosity Objects discovered by \emph{Spitzer} in cores previously believed to be starless, look very similar to other cores with more luminous protostars, indicating that 350 $\mu$m observations of these objects may be a key component in confirming their status as very low luminosity, embedded objects.  Future work will concentrate on expanding the sample of cores observed with SHARC-II and combining this dataset with others being assembed at other submillimeter and millimeter wavelengths in order to assemble a more complete picture of the processes involved in low-mass star formation.

The authors extend our gratitude to Paul Ho, the referee, as well as Jes J\o rgensen and Yancy Shirley, for insightful comments that have led to a much improved paper.  We also acknowledge the support and assistance provided by the SHARC-II at Caltech: Colin Borys, Darren Dowell, Attila Kovacs, and their colleagues.  We thank the staff at the CSO, as well as Katelyn Allers, Jo-hsin Chen, Jackie Kessler-Silacci, and Claudia Knez, for their assistance with obtaining the observations.  We acknowledge the data analysis facilities provided by the Starlink Project which is run by CCLRC on behalf of PPARC.  We thank the Lorentz Center in Leiden for hosting several meetings that contributed to this paper.  Support for this work, part of the Spitzer Legacy Science Program, was provided by NASA through contract 1224608 issued by the Jet Propulsion Laboratory, California Institute of Technology, under NASA contract 1407.  This work was also supported by NASA Origins grant NNG04GG24G.


\clearpage
\begin{deluxetable}{lcccccccc}
\tabletypesize{\scriptsize}
\tablewidth{0pt}
\tablecaption{\label{info}Observing Information}
\tablehead{
\colhead{CORE\tablenotemark{a}}            & \colhead{Map center}  & \colhead{Map center}  &
\colhead{Dist}          & \colhead{Ref\tablenotemark{b}}  &          \colhead{Project\tablenotemark{c}}  & 
\colhead{Obs.Date}       & \colhead{1$\sigma$ noise} 	& \colhead{$\sigma_{M}$\tablenotemark{d}}\\ &
\colhead{RA (2000.0)} &
\colhead{Dec (2000.0)}      &\colhead{(pc)} &
\colhead{}         &  \colhead{}         & \colhead{(mm/yyyy)}&
\colhead{(mJy beam$^{-1}$)}	& \colhead{(\msun\ beam$^{-1}$)}}
\startdata
L1455                     & 03 27 41.0 & +30 12 45.0 & 250& 1  & Per     & 11/2005 & 140 & 6.49(-3) \\
IRAS 03282+3035           & 03 31 20.4 & +30 45 24.7 & 250& 1  & Per     & 11/2005 & 150 & 6.96(-3) \\
IRAS 03292+3039           & 03 32 18.0 & +30 49 47.0 & 250& 1  & Per     & 11/2005 & 190 & 8.81(-3) \\
BOLO68                    & 03 32 28.1 & +31 02 17.5 & 250& 1  & Per     & 11/2005 & 55  & 2.55(-3) \\
IRAS 03301+3111           & 03 33 12.9 & +31 21 24.2 & 250& 1  & Per/IRS & 09/2004 & 160 & 7.42(-3) \\
LkH$\alpha$ 327           & 03 33 30.4 & +31 10 50.4 & 250& 1  & Per/IRS & 09/2004 & 90  & 4.17(-3) \\
HH211-MM                  & 03 43 56.8 & +32 00 50.2 & 250& 1  & Per     & 11/2005 & 290 & 1.34(-2) \\
IRAS 03439+3233           & 03 47 05.5 & +32 43 08.5 & 250& 1  & Per/IRS & 09/2004 & 270 & 1.25(-2) \\
IRAS 03445+3242           & 03 47 41.6 & +32 51 43.8 & 250& 1  & Per/IRS & 09/2004 & 220 & 1.02(-2) \\
IRAM 04191+1522           & 04 21 56.9 & +15 29 45.0 & 140& 2  & c2d     & 11/2005 & 100 & 1.45(-3) \\
L1521B                    & 04 24 14.9 & +26 36 53.0 & 140& 2  & Oth     & 11/2005 & 35  & 5.09(-4) \\
L1521F                    & 04 28 38.9 & +26 51 35.0 & 140& 2  & c2d     & 11/2005 & 70  & 1.02(-3) \\
L1521E                    & 04 29 13.6 & +26 14 05.0 & 140& 2  & c2d+    & 11/2005 & 70  & 1.02(-3) \\
B18-1 (TMC-2A)            & 04 31 57.7 & +24 32 30.0 & 140& 2  & c2d     & 11/2005 & 75  & 1.09(-3) \\
TMC2 (L1529)              & 04 32 44.8 & +24 23 58.0 & 140& 2  & c2d     & 11/2005 & 45  & 6.54(-4) \\
B18-4 (TMC3)              & 04 35 37.5 & +24 09 20.0 & 140& 2  & GTO     & 11/2005 & 150 & 2.18(-3) \\
TMC1-A                    & 04 41 18.9 & +25 48 45.0 & 140& 2  & c2d     & 11/2005 & 35  & 5.09(-4) \\
L1544                     & 05 04 16.6 & +25 10 48.0 & 140& 2  & GTO     & 11/2005 & 30  & 4.36(-4) \\
L1582B (RNO43)            & 05 32 19.4 & +12 49 43.0 & 400& 3  & exc     & 10/2003 & 130 & 1.54(-2) \\
B35A (L1594)              & 05 44 29.2 & +09 08 52.0 & 400& 3  & c2d     & 11/2005 & 220 & 2.61(-2) \\
DC255.4-3.9 (BHR16)       & 08 05 34.0 & -39 09 12.0 & 440& 4  & c2d     & 11/2005 & 200 & 2.87(-2) \\
CG30 (DC253.3-1.6, BHR12) & 08 09 32.7 & -36 04 58.0 & 450& 5  & c2d     & 11/2005 & 380 & 5.71(-2) \\
L134A                     & 15 53 36.3 & -04 35 25.9 & 110& 6  & c2d     & 03/2005 & 55  & 4.94(-4) \\
L43 (RNO91)               & 16 34 33.0 & -15 47 08.0 & 125& 7  & c2d     & 03,06/2005 & 130 & 1.51(-3) \\
L146 (CB68)               & 16 57 20.5 & -16 09 02.0 & 125& 7  & c2d     & 06/2004 & 360 & 4.17(-3) \\
B59 (L1746)               & 17 11 22.7 & -27 24 28.0 & 125& 7  & c2d     & 06/2005 & 400 & 4.64(-3) \\
L492 (CB128)              & 18 15 48.4 & -03 45 47.0 & 270& 8  & c2d     & 06/2005 & 60  & 3.25(-3) \\
L507 (CB130)              & 18 16 16.4 & -02 32 38.0 & 270& 8  & c2d     & 06/2005 & 35  & 1.89(-3) \\
L328 (CB131)              & 18 16 59.5 & -18 02 30.0 & 270& 8  & c2d     & 06/2005 & 80  & 4.33(-3) \\
L483                      & 18 17 29.9 & -04 39 40.0 & 270& 8  & c2d     & 06/2005 & 280 & 1.51(-2) \\
EC74                      & 18 29 55.7 & +01 14 31.6 & 260& 9  & IRS     & 06/2004 & 510 & 2.56(-2) \\
EC88                      & 18 29 57.6 & +01 13 00.6 & 260& 9  & IRS     & 06/2004 & 700 & 3.51(-2) \\
L723                      & 19 17 53.8 & +19 12 18.5 & 300& 10 & c2d     & 05/2003 & 80  & 5.34(-3) \\
CB188 (L673-1)            & 19 20 15.0 & +11 36 08.0 & 300& 11 & c2d     & 06/2004 & 60  & 4.01(-3) \\
B335                      & 19 37 01.1 & +07 34 10.8 & 230& 12 & GTO     & 05/2003 & 340 & 1.33(-2) \\
L694-2 (B143, CB200)      & 19 41 04.3 & +10 57 09.0 & 230& 12 & c2d     & 11/2005 & 90  & 3.53(-3) \\
L1152-1                   & 20 35 46.4 & +67 53 02.0 & 325& 13 & c2d     & 11/2005 & 80  & 6.27(-3) \\
L1152-2                   & 20 36 20.2 & +67 56 33.0 & 325& 13 & c2d     & 11/2005 & 75  & 5.88(-3) \\
L1041-2                   & 20 37 20.7 & +57 44 13.0 & 400& 14 & exc     & 06/2004 & 300 & 3.56(-2) \\
L1157                     & 20 39 06.2 & +68 02 16.0 & 325& 13 & exc     & 11/2005 & 510 & 4.00(-2) \\
L1148B                    & 20 40 56.8 & +67 23 05.5 & 325& 13 & c2d     & 06/2005 & 80  & 6.27(-3) \\
L1228                     & 20 57 11.8 & +77 35 47.9 & 200& 15 & c2d     & 09/2004 & 960 & 2.85(-2) \\
Bern48 (RNO129)           & 20 59 15.0 & +78 22 59.9 & 200& 15 & c2d     & 05/2003 & 350 & 1.04(-2) \\
L1177 (CB230)             & 21 17 40.0 & +68 17 31.9 & 288& 13 & exc     & 05/2003 & 230 & 1.42(-2) \\
L1021                     & 21 21 47.0 & +50 58 16.0 & 250& 16 & c2d     & 09/2004 & 110 & 5.10(-3) \\
L1014                     & 21 24 07.0 & +49 59 09.0 & 250& 16 & c2d     & 09/2004 & 100 & 4.64(-3) \\
L1165                     & 22 06 50.4 & +59 02 46.0 & 300& 14 & c2d     & 06/2005 & 90  & 6.01(-3) \\
L1221                     & 22 28 04.7 & +69 00 57.0 & 250& 17 & c2d     & 09/2004 & 190 & 8.81(-3) \\
L1251A-1                  & 22 30 32.2 & +75 14 09.4 & 300& 18 & c2d     & 11/2005 & 170 & 1.14(-2) \\
L1251A-2                  & 22 31 03.3 & +75 13 39.0 & 300& 18 & c2d     & 11/2005 & 180 & 1.20(-2) \\
L1251C (L1251N)           & 22 35 24.1 & +75 17 07.9 & 300& 18 & c2d     & 09/2003 & 140 & 9.35(-3) \\
L1251B (L1251E)           & 22 38 47.1 & +75 11 28.8 & 300& 18 & c2d     & 05/2003 & 220 & 1.47(-2) \\
\hline						    
L507 (CB130)$^{450}$      & 18 16 16.4 & -02 32 38.0 & 270& 8  & c2d     & 06/2005 & 55  & 2.98(-3) \\
L483$^{450}$              & 18 17 29.9 & -04 39 40.0 & 270& 8  & c2d     & 06/2005 & 150 & 8.11(-3) \\
L673-7$^{450}$            & 19 21 34.8 & +11 21 24.0 & 300& 11 & c2d     & 06/2005 & 30  & 2.00(-3) \\
\enddata\\
\tablenotetext{a}{A core with superscript ``450'' indicates that the map
of that core was observed at 450 $\mu$m.}
\tablenotetext{b}{References  (1) Enoch et al. (2006); (2) Kenyon, Dobrzycka, \& Hartmann (1994); (3) Murdin \& Penston (1977); (4) Herbst (1975); (5) Woermann, Gaylard, \& Otrupcek (2001); (6) Franco (1989); (7) de Geus et al. (1989); (8) Strai{\v z}ys et al. (2003); (9) Strai{\v z}ys et al. (1996); (10) Goldsmith et al. (1984); (11) Herbig \& Jones (1983); (12) Kawamura et al. (2001); (13) Strai{\v z}ys et al. (1992); (14) Dobashi et al. (1994); (15) Kun (1998); (16) Pagani \& Breart de Boisanger (1996); (17) Yonekura et al. (1997); (18) Kun \& Prusti (1993)}
\tablenotetext{c}{The project from which the core was selected \\
Per:  Observed in Perseus c2d cloud map\\
IRS:  Observed in c2d IRS program\\
c2d:  Observed in cores c2d program\\
Oth:  Selected for SHARC-II observations for other reasons\\
c2d+:  Observed in c2d-related GO program\\
GTO:  Observed in GTO observations\\
exc:  Originally part of c2d cores program but cut from final target list}
\tablenotetext{d}{Distance-independent 1$\sigma$ map sensitivity in units of \msun\ beam$^{-1}$.  An entry in this column of X.XX(-Y) should be interpreted as X.XX\ee{-Y}.}

\end{deluxetable}
\begin{deluxetable}{lccccc}
\tabletypesize{\scriptsize}
\tablewidth{0pt}
\tablecaption{\label{overlap}Overlap with Complementary Programs}
\tablehead{
\colhead{CORE\tablenotemark{a}} &\colhead {SCUBA\tablenotemark{b}} & \colhead{MAMBO\tablenotemark{c}} & \colhead{SIMBA\tablenotemark{d}} &	\colhead{FCRAO\tablenotemark{e}}	& \colhead{SCUBA Starless\tablenotemark{f}}}
\startdata
L1455	&	N	&	N	&	N	&	N	&	N\\
IRAS 03282+3035	&	N	&	N	&	N	&	N	&	N\\
IRAS 03292+3039	&	N	&	N	&	N	&	N	&	N\\
BOLO68	&	N	&	N	&	N	&	N	&	N\\
IRAS 03301+3111	&	N	&	N	&	N	&	N	&	N\\
LkH$\alpha$ 327	&	N	&	N	&	N	&	N	&	N\\
HH211-MM	&	N	&	N	&	N	&	N	&	N\\
IRAS 03439+3233	&	N	&	N	&	N	&	N	&	N\\
IRAS 03445+3242	&	N	&	N	&	N	&	N	&	N\\
IRAM 04191+1522	&	Y	&	N	&	Y	&	N	&	N\\
L1521B	&	N	&	N	&	N	&	N	&	N\\
L1521F	&	Y	&	Y	&	N	&	N	&	N\\
L1521E	&	N	&	N	&	N	&	N	&	Y\\
B18-1 (TMC-2A)	&	Y	&	Y	&	N	&	Y	&	N\\
TMC2 (L1529)	&	N	&	Y	&	N	&	N	&	N\\
B18-4 (TMC3)	&	N	&	N	&	N	&	Y	&	N\\
TMC1-A	&	Y	&	Y	&	N	&	Y	&	N\\
L1544	&	N	&	N	&	N	&	N	&	Y\\
L1582B (RNO43)	&	Y	&	N	&	N	&	Y	&	N\\
B35A (L1594)	&	Y	&	N	&	Y	&	N	&	N\\
DC255.4-3.9 (BHR16)	&	N	&	N	&	Y	&	N	&	N\\
CG30 (DC253.3-1.6, BHR12)	&	N	&	N	&	Y	&	N	&	N\\
L134A	&	N	&	N	&	N	&	N	&	N\\
L43 (RNO91)	&	Y	&	N	&	N	&	N	&	Y\\
L146 (CB68)	&	Y	&	N	&	N	&	Y	&	N\\
B59 (L1746)	&	N	&	N	&	N	&	Y	&	N\\
L492 (CB128)	&	Y	&	Y	&	N	&	N	&	Y\\
L507 (CB130)	&	N	&	N	&	Y	&	Y	&	Y\\
L328 (CB131)	&	N	&	N	&	N	&	N	&	Y\\
L483	&	N	&	N	&	Y	&	Y	&	N\\
EC74	&	N	&	N	&	N	&	N	&	N\\
EC88	&	N	&	N	&	N	&	N	&	N\\
L723	&	Y	&	N	&	N	&	Y	&	N\\
CB188 (L673-1)	&	N	&	Y	&	N	&	Y	&	N\\
B335	&	N	&	N	&	N	&	N	&	N\\
L694-2 (B143, CB200)	&	N	&	N	&	N	&	N	&	Y\\
L1152-1	&	Y	&	N	&	N	&	N	&	N\\
L1152-2	&	Y	&	N	&	N	&	N	&	N\\
L1041-2	&	N	&	Y	&	N	&	Y	&	N\\
L1157	&	Y	&	N	&	N	&	Y	&	N\\
L1148B	&	N	&	Y	&	N	&	N	&	N\\
L1228	&	Y	&	Y	&	N	&	N	&	N\\
Bern48 (RNO129)	&	N	&	Y	&	N	&	N	&	N\\
L1177 (CB230)	&	Y	&	Y	&	N	&	Y	&	N\\
L1021	&	N	&	Y	&	N	&	Y	&	N\\
L1014	&	N	&	Y	&	N	&	Y	&	N\\
L1165	&	N	&	N	&	N	&	Y	&	N\\
L1221	&	Y	&	N	&	N	&	Y	&	N\\
L1251A-1	&	N	&	Y	&	N	&	N	&	N\\
L1251A-2	&	N	&	Y	&	N	&	N	&	N\\
L1251C (L1251N)	&	Y	&	N	&	N	&	N	&	N\\
L1251B (L1251E)	&	Y	&	N	&	N	&	N	&	N\\
\hline
L507 (CB130)$^{450}$ &	N	&	N	&	N	&	N	&	N\\
L483$^{450}$	&	N	&	N	&	N	&	N	&	N\\
L673-7$^{450}$	&	N	&	Y	&	N	&	Y	&	N\\
\hline
\\
\hline
Total\tablenotemark{g} & 18 & 16 & 6 & 18 & 7\\
\enddata\\
\tablenotetext{a}{A core with superscript ``450'' indicates that the map of that core was observed at 450 $\mu$m.}
\tablenotetext{b}{Indicates whether or not the core is included in an 850 \um\ SCUBA survey by Young et al. (2006).}
\tablenotetext{c}{Indicates whether or not the core is included in a 1.2 mm MAMBO survey by J. Kauffmann et al. (2006, in preparation).}
\tablenotetext{d}{Indicates whether or not the core is included in a 1.2 mm SIMBA survey by K. Brede et al. (2006, in preparation).}
\tablenotetext{e}{Indicates whether or not the core is included in an \nthp\ (1-0) and CS (2-1) FCRAO survey by C. De Vries et al. (2006, in preparation).}
\tablenotetext{f}{Indicates whether or not the core is included in a project to construct radiative transfer models of starless cores using 450 and 850 \um\ data from the SCUBA archive by Y. Shirley et al. (2006, in preparation).}
\tablenotetext{g}{Total number of the cores presented in this survey that are also included in each of the complementary programs.}

\end{deluxetable}
\begin{table}
\caption{\label{caltable}Calibrators}  
\begin{tabular}{lcclccc}
\hline
Date   &    Calibrator & Filter & $\tau_{225\ GHz}$ & C$_{beam}$& C$_{20}$ &C$_{40}$\\
       &               & (\um)  &                   & (Jy beam$^{-1}$ nV$^{-1}$)  &    (Jy nV$^{-1}$) &    (Jy nV$^{-1}$) \\
\hline
05/17/03  &  Mars      &   350  &   0.07  & 5.82  & 0.22 & 0.19 \\
05/18/03  &  Uranus    &   350  &   0.085 & 6.83  & 0.22 & 0.21 \\
\hline  
09/29/03  &  Uranus    &   350  &   0.054 & 5.75  & 0.20 & 0.19 \\
09/30/03  &  Uranus    &   350  &   0.09  & 9.43  & 0.26 & 0.25 \\
10/01/03  &  Uranus    &   350  &   0.063 & 6.23  & 0.21 & 0.21 \\
\hline
06/19/04  &  Neptune   &   350  &   0.08  & 6.71  & 0.21 & 0.19 \\
          &  Uranus    &   350  &   0.09  & 8.45  & 0.24 & 0.22 \\
\hline
09/24/04  &  Uranus    &   350  &   0.065 & 7.55  & 0.23 & 0.21 \\
09/25/04  &  Uranus    &   350  &   0.081 & 7.14  & 0.22 & 0.21 \\
          &  Neptune   &   350  &   0.071 & 6.43  & 0.19 & 0.18 \\
          &  Uranus    &   350  &   0.061 & 5.57  & 0.16 & 0.17 \\
09/28/04  &  Neptune   &   350  &   0.065 & 10.38 & 0.24 & 0.19 \\
\hline
06/16/05  &  Neptune   &   350  &   0.045 & 6.94  & 0.22 & 0.17 \\
          &  Uranus    &   350  &   0.045 & 6.86  & 0.23 & 0.18 \\
          &  Uranus    &   350  &   0.039 & 6.42  & 0.32 & 0.17 \\
06/17/05  &  Neptune   &   350  &   0.048 & 6.29  & 0.22 & 0.19 \\
          &  Uranus    &   350  &   0.049 & 6.36  & 0.23 & 0.19 \\
\hline
11/03/05  &  Uranus    &   350  &   0.061 & 6.42  & 0.20 & 0.16 \\
          &  Uranus    &   350  &   0.055 & 5.98  & 0.18 & 0.15 \\
          &  Uranus    &   350  &   0.070 & 5.35  & 0.16 & 0.14 \\
11/04/05  &  Uranus    &   350  &   0.071 & 5.72  & 0.18 & 0.14 \\
          &  Uranus    &   350  &   0.082 & 7.05  & 0.22 & 0.20 \\
          &  Uranus    &   350  &   0.062 & 7.54  & 0.21 & 0.18 \\
11/05/05  &  Uranus    &   350  &   0.059 & 6.38  & 0.22 & 0.19 \\
          &  Uranus    &   350  &   0.065 & 8.29  & 0.26 & 0.22 \\
          &  Uranus    &   350  &   0.056 & 5.43  & 0.18 & 0.17 \\
11/12/05  &  Uranus    &   350  &   0.077 & 7.09  & 0.21 & 0.16 \\
          &  Uranus    &   350  &   0.087 & 7.45  & 0.22 & 0.17 \\
          &  Uranus    &   350  &   0.076 & 6.62  & 0.20 & 0.15 \\
\hline
\\\hline   	                         	    	   
06/15/05  &  Neptune   &   450  &   0.058 & 10.75 & 0.23 & 0.20 \\
          &  Neptune   &   450  &   0.058 & 10.93 & 0.24 & 0.19 \\
          &  Neptune   &   450  &   0.058 & 11.86 & 0.37 & 0.24 \\
          &  Uranus    &   450  &   0.063 & 9.71  & 0.22 & 0.18 \\
\hline
\end{tabular}\\
\end{table}
\begin{table}
\caption{\label{caltable2}Average Calibration Factors}  
\begin{tabular}{lcccc}
\hline
Date & Filter & Average C$_{beam}$ & Average C$_{20}$ & Average C$_{40}$   \\
     &($\mu$m)&  (Jy beam$^{-1}$ nV$^{-1}$)  &    (Jy nV$^{-1}$) &    (Jy nV$^{-1}$)     \\
\hline
05/2003        & 350 & $6.33  \pm 0.51$ & $0.22 \pm 0.00$ & $0.20 \pm 0.01$\\
09/2003        & 350 & $7.14  \pm 1.63$ & $0.22 \pm 0.03$ & $0.22 \pm 0.02$\\
06/2004        & 350 & $7.58  \pm 0.87$ & $0.23 \pm 0.02$ & $0.21 \pm 0.02$\\
09/2004        & 350 & $7.41  \pm 1.63$ & $0.21 \pm 0.03$ & $0.19 \pm 0.02$\\
06/2005        & 350 & $6.57  \pm 0.27$ & $0.24 \pm 0.04$ & $0.18 \pm 0.01$\\
11/2005        & 350 & $6.61  \pm 0.87$ & $0.20 \pm 0.03$ & $0.17 \pm 0.02$\\
06/2005        & 450 & $10.81 \pm 0.76$ & $0.27 \pm 0.06$ & $0.20 \pm 0.02$\\
\hline \\ \hline
All Runs$^{*}$ & 350 & $6.88  \pm 1.16$ & $0.22 \pm 0.04$ & $0.19 \pm 0.03$\\
\hline
\end{tabular}\\

$^{*}$ The average flux conversion factor over all the runs is only presented for 350 $\mu$m as the June 2005 run was the only run in which 450 $\mu$m data was taken.
\end{table}
\begin{deluxetable}{lccrrccccc}
\tabletypesize{\scriptsize}
\tablewidth{0pt}
\tablecaption{\label{properties1}Source Properties}
\tablehead{
\colhead{SOURCE\tablenotemark{a}}        &
\colhead{Barycenter}                     &
\colhead{Barycenter}                     &
\colhead{S$_{20}$\tablenotemark{b}}      &
\colhead{S$_{40}$\tablenotemark{b}}      &
\colhead{Peak}                           &
\colhead{Peak}                           &
\colhead{$\delta_{pk}$\tablenotemark{c}} &
\colhead{Peak}                           &
\colhead{SST\tablenotemark{d}}           \\
\colhead{}                               &
\colhead{RA}                             &
\colhead{Dec}                            &
\colhead{}                               &
\colhead{}                               &
\colhead{RA}                             &
\colhead{Dec}                            &
\colhead{}                               &
\colhead{Flux}                           &
\colhead{}                               \\
\colhead{}                               &
\colhead{(J2000)}                        &
\colhead{(J2000)}                        &
\colhead{(Jy)}                           &
\colhead{(Jy)}                           &
\colhead{(J2000)}                        &
\colhead{(J2000)}                        &
\colhead{(\as)}                          &
\colhead{(Jy beam$^{-1}$)}               &
\colhead{}                    
}               
\startdata
L1455-IRS5       &  03 27 38.39 & +30 14 01.5 & 4.3 (0.6)  & 9.8 (1.5)  & 03 27 38.32 & +30 13 58.6 & 3.0 & 1.6 (0.2)  & Y \\
L1455-IRS1       &  03 27 39.21 & +30 13 03.8 & 11.6 (1.9) & 13.0 (2.1) & 03 27 39.19 & +30 13 03.6 & 0.3 & 7.6 (1.1)  & Y \\
L1455-IRS4       &  03 27 43.26 & +30 12 29.4 & 5.3 (0.8)  & 6.0 (1.0)  & 03 27 43.31 & +30 12 29.6 & 0.7 & 3.1 (0.5)  & Y \\
L1455-IRS2       &  03 27 48.44 & +30 12 09.6 & 3.0 (0.5)  & 5.1 (0.8)  & 03 27 48.49 & +30 12 10.2 & 0.9 & 1.2 (0.2)  & Y\tablenotemark{e} \\
IRAS 03282+3035  &  03 31 21.10 & +30 45 30.2 & 11.1 (1.8) & 12.3 (2.1) & 03 31 21.12 & +30 45 30.4 & 0.3 & 8.2 (1.2)  & Y \\
IRAS 03292+3039  &  03 32 18.03 & +30 49 46.5 & 14.1 (2.3) & 17.5 (2.8) & 03 32 18.06 & +30 49 46.2 & 0.5 & 10.4 (1.6) & Y \\
BOLO68           &  03 32 29.32 & +31 02 39.6 & 1.7 (0.3)  & 2.9 (0.4)  & 03 32 29.30 & +31 02 41.0 & 1.4 & 0.8 (0.1)  & Y \\
IRAS 03301+3111  &      -       &      -      & 0.19 (0.06)&    -       &       -     &       -     & -   & 0.6 (0.1) & Y \\
LkH$\alpha$ 327  &      -       &      -      &($<$ 0.09)  &($<$ 0.19)  &       -     &       -     & -   &  -   & Y \\
HH211-MM         &  03 43 56.93 & +32 00 48.9 & 16.4 (2.6) & 25.2 (4.1) & 03 43 56.94 & +32 00 49.4 & 0.5 & 10.4 (1.6) & Y\tablenotemark{f} \\
IRAS 03439+3233  &      -       &      -      & 1.1 (0.2)  &    -       &       -     &       -     & -   & 0.6 (0.1) & Y \\
IRAS 03445+3242  &  03 47 41.77 & +32 51 47.8 & 6.4 (1.0)  & 11.5 (1.7) & 03 47 41.83 & +32 51 48.3 & 0.9 & 3.6 (0.5) & Y \\
IRAM 04191+1522\tablenotemark{g} &  04 21 57.04 & +15 29 49.1 & 3.6 (0.5)  & 5.0 (0.8)  & 04 21 57.07 & +15 29 47.4 & 1.8 & 1.6 (0.2) & Y \\
IRAS 04191+1523\tablenotemark{h} &  04 22 00.60 & +15 30 23.4 & 3.7 (0.6)  &	 -      & 04 22 00.43 & +15 30 24.7 & 2.8 & 2.0 (0.3) & Y \\
L1521B           &       -      &      -      & ($<$ 0.03) & ($<$ 0.06) &       -     &        -    & -   &   -  & U \\
L1521F\tablenotemark{g} &  04 28 39.11 & +26 51 36.4 & 2.8 (0.4)  & 4.5 (0.7)  & 04 28 38.96 & +26 51 35.8 & 2.1 & 1.3 (0.2) & Y \\
L1521E           &       -      &      -      &($<$ 0.02)  &($<$ 0.08)  &       -     &        -    & -   &   -  & U \\
B18-1            &       -      &      -      &($<$ 0.05)  &($<$ 0.13)  &       -     &        -    & -   &   -  & N \\
TMC2             &       -      &      -      &($<$ 0.03)  &($<$ 0.07)  &       -     &        -    & -   &   -  & N \\
B18-4            &  04 35 37.23 & +24 09 15.8 & 1.9 (0.3)  & 4.7 (0.7)  & 04 35 37.68 & +24 09 19.2 & 7.0 & 1.6 (0.2) & U \\
IRAS 04325+2402\tablenotemark{i}  &  04 35 35.27 & +24 08 20.3 & 3.8 (0.6)  & 7.1 (1.1)  & 04 35 35.31 & +24 08 22.6 & 2.4 & 0.6 (0.1) & U \\
TMC1-A           &       -      &      -      &($<$ 0.03)  &($<$ 0.07)  &       -     &        -    & -   &   -  & N \\
L1544            &       -      &      -      &($<$ 0.03)  &($<$ 0.05)  &       -     &        -    & -   &   -  & U \\
L1582B           &  05 32 19.34 & +12 49 41.1 & 12.2 (2.0) & 14.2 (2.4) & 05 32 19.30 & +12 49 40.5 & 0.8 & 9.1 (1.4) & U \\
B35A-SMM1        &  05 44 29.49 & +09 08 54.7 & 11.8 (1.8) & 17.3 (2.6) & 05 44 29.42 & +09 08 54.4 & 1.1 & 6.3 (1.0) & Y \\
B35A-SMM2        &  05 44 30.41 & +09 09 15.7 & 5.0 (0.8)  &   -        & 05 44 30.29 & +09 09 16.3 & 1.9 & 2.0 (0.3) & N \\
B35A-SMM3        &  05 44 31.68 & +09 09 01.1 & 3.6 (0.5)  &   -        & 05 44 31.63 & +09 09 00.9 & 0.8 & 1.5 (0.2) & Y \\
DC255.4-3.9\tablenotemark{j,k}      &       -      &      -      &($<$ 0.08)  &($<$ 0.21)  &       -     &        -    & -   &   -  & Y \\
CG30-SMM1\tablenotemark{j} &  08 09 32.55 & -36 05 16.6 & 16.3 (2.6) & 24.7 (4.0) & 08 09 32.50 & -36 05 16.6 & 0.6 & 10.2 (1.5) & Y \\
CG30-SMM2        &  08 09 32.87 & -36 04 56.3 & 15.6 (2.5) & 25.9 (4.1) & 08 09 32.90 & -36 04 55.6 & 0.8 & 9.4 (1.4) & Y \\
L134A            &       -      &      -      &($<$ 0.05)  &($<$ 0.09)  &       -     &        -    & -   &   -  & N \\
L43-RNO91        &  16 34 29.42 & -15 47 01.3 & 6.9 (1.0)  & 9.9 (1.6)  & 16 34 29.29 & -15 47 01.4 & 1.9 & 3.0 (0.5) & Y \\
L43-SMM1         &  16 34 32.61 & -15 46 32.9 & 0.9 (0.3)  & 1.0 (0.4)  & 16 34 32.68 & -15 46 32.7 & 1.0 & 0.4 (0.1) & N \\
L43-SMM2         &  16 34 35.08 & -15 47 03.2 & 3.0 (0.7)  & 6.5 (1.3)  & 16 34 35.06 & -15 47 03.5 & 0.4 & 1.0 (0.2) & N \\
L43-SMM3         &  16 34 37.15 & -15 47 37.4 & 2.2 (0.6)  & 4.6 (1.0)  & 16 34 37.11 & -15 47 34.2 & 3.2 & 0.8 (0.1) & Y\tablenotemark{l} \\
L146             &  16 57 19.73 & -16 09 20.9 & 9.9 (1.5)  & 19.1 (2.9) & 16 57 19.77 & -16 09 22.2 & 1.4 & 4.8 (0.7) & Y \\
B59\tablenotemark{m} &  17 11 23.16 & -27 24 32.8 & 40.2 (7.1) & 45.2 (8.0) & 17 11 23.18 & -27 24 32.9 & 0.3 & 30.3 (5.5) & Y \\
L492             &  18 15 47.65 & -03 45 48.6 & 0.8 (0.3)  & 2.2 (0.6)  & 18 15 48.23 & -03 45 49.4 & 8.7 & 0.3 (0.1) & N \\
L507\tablenotemark{j} &  18 16 16.56 & -02 32 40.7 & 1.6 (0.4)  & 2.8 (0.7)  & 18 16 16.57 & -02 32 40.4 & 0.3 & 0.7 (0.1) & Y \\
L328-SMM1        &  18 16 59.50 & -18 02 09.5 & 3.1 (0.7)  & 6.2 (1.2)  & 18 16 59.55 & -18 02 06.5 & 3.1 & 1.1 (0.2) & N \\
L328-SMM2\tablenotemark{j} &      -       &      -      & 2.0 (0.5) & -  & 18 16 59.69 & -18 02 31.6 & - & 0.9 (0.2) & Y \\
L328-SMM3        &  18 17 00.86 & -18 02 08.8 & 2.0 (0.5)  &     -      & 18 17 00.88 & -18 02 09.0 & 0.4 & 0.7 (0.1) & N \\
L483             &  18 17 29.90 & -04 39 40.8 & 30.3 (4.8) & 50.1 (7.8) & 18 17 30.07 & -04 39 42.4 & 3.1 & 11.8 (2.1) & Y \\
EC74             &      -       &      -      &  3.0 (0.5) &     -      &       -     &       -     & -   &  1.8 (0.3) & Y \\
EC88             &      -       &      -      &  2.2 (0.4) &     -      &       -     &       -     & -   &  1.6 (0.2) & Y \\
L723             &  19 17 53.41 & +19 12 12.8 & 8.5 (1.4)  & 11.3 (1.8) & 19 17 53.37 & +19 12 12.9 & 0.5 & 4.7 (0.7) & Y \\
CB188            &  19 20 15.27 & +11 35 48.7 & 1.0 (0.2)  & 1.0 (0.2)  & 19 20 15.31 & +11 35 47.8 & 1.0 & 0.6 (0.1) & Y \\
B335             &  19 37 01.20 & +07 34 10.2 & 20.7 (3.3) & 25.9 (4.1) & 19 37 01.20 & +07 34 10.0 & 0.2 & 11.4 (1.7) & Y \\
L694-2           &      -       &      -      &($<$ 0.11)  &($<$ 0.20)  &       -     &       -     & -   &   -  & N \\
L1152-1          &  20 35 46.75 & +67 53 02.1 & 4.1 (0.7)  & 7.3 (1.1)  & 20 35 46.83 & +67 53 02.8 & 0.8 & 2.5 (0.4) & Y \\
L1152-2          &  20 36 20.20 & +67 56 31.8 & 0.6 (0.1)  &	 -      & 20 36 20.27 & +67 56 32.1 & 0.5 & 0.7 (0.1) & Y \\
L1041-2          &  20 37 21.34 & +57 44 15.4 & 8.2 (1.2)  & 15.8 (2.4) & 20 37 21.21 & +57 44 15.0 & 1.1 & 4.0 (0.6) & U \\
L1157            &  20 39 06.82 & +68 02 22.2 & 23.5 (3.5) & 36.4 (5.8) & 20 39 06.95 & +68 02 21.7 & 0.9 & 12.6 (1.9) & U \\
L1148B\tablenotemark{j} &      -       &      -      &($<$ 0.04)  &($<$ 0.11)  &       -     &       -     & - &   -  & Y \\
L1228            &  20 57 16.52 & +77 35 42.2 & 15.5 (2.5) & 21.7 (3.5) & 20 57 16.08 & +77 35 43.9 & 2.2 & 11.9 (1.8) & Y \\
Bern48           &  20 59 16.57 & +78 23 07.8 & 21.7 (3.5) & 21.4 (3.4) & 20 59 16.50 & +78 23 08.2 & 0.5 & 13.2 (2.0) & Y \\
L1177            &  21 17 39.78 & +68 17 33.1 & 12.8 (2.0) & 18.0 (2.9) & 21 17 39.76 & +68 17 32.8 & 0.3 & 7.0 (1.0) & U \\
L1021            &      -       &      -      &($<$ 0.12)  & ($<$ 0.23) &       -     &       -     & -   &   -  & N \\
L1014\tablenotemark{g} &  21 24 08.16 & +49 59 09.3 & 1.5 (0.2)  & 3.2 (0.5)  & 21 24 08.27 & +49 59 08.7 & 1.2 & 0.8 (0.1) & Y \\
L1165            &  22 06 51.03 & +59 02 40.3 & 14.3 (2.6) & 17.0 (2.9) & 22 06 51.13 & +59 02 40.3 & 0.8 & 8.5 (1.5) & Y \\
L1221-SMM1       &  22 28 03.48 & +69 01 21.9 & 4.6 (0.7)  & 9.3 (1.4)  & 22 28 03.35 & +69 01 22.9 & 1.2 & 2.5 (0.4) & Y \\
L1221-SMM2\tablenotemark{j} &  22 28 07.99 & +69 00 42.8 & 4.9 (0.8)  & 6.4 (1.0)  & 22 28 07.86 & +69 00 43.2 & 0.8 & 3.6 (0.5) & Y \\
L1251A-1         &  22 30 32.88 & +75 14 12.9 & 5.6 (0.8)  & 9.4 (1.4)  & 22 30 32.83 & +75 14 13.5 & 0.6 & 3.0 (0.5) & Y\tablenotemark{n} \\
L1251A-2         &  22 31 06.06 & +75 13 40.9 & 3.8 (0.6)  & 7.5 (1.1)  & 22 31 06.05 & +75 13 41.4 & 0.5 & 1.9 (0.3) & Y\tablenotemark{n} \\
L1251C           &  22 35 25.02 & +75 17 06.2 & 10.4 (1.6) & 20.8 (3.2) & 22 35 24.22 & +75 17 08.9 & 4.1 & 5.4 (0.8) & Y \\
L1251B-SMM1      &	-       &    -	      &  -         &    -       & 22 38 46.84 & +75 11 39.5 & -   & 6.2 (0.9) & Y \\
L1251B-SMM2      &  22 38 48.64 & +75 11 36.5 & 20.0 (3.0) & 36.9 (5.5) & 22 38 49.58 & +75 11 35.4 & 3.8 & 6.5 (1.0) & N \\
\hline
L507$^{450}$     &  18 16 16.58 & -02 32 39.9 & 0.8 (0.3)  & 1.1 (0.4)  & 18 16 16.68 & -02 32 38.8 & 1.8 & 0.5 (0.1) & Y \\
L483$^{450}$     &  18 17 29.97 & -04 39 37.2 & 11.4 (2.0) & 20.0 (3.3) & 18 17 30.20 & -04 39 37.6 & 3.4 & 6.5 (1.2) & Y \\
L673-7$^{450}$\tablenotemark{j} &  19 21 34.92 & +11 21 20.1 & 0.8 (0.3)  & 1.1 (0.4)  & 19 21 34.88 & +11 21 20.0 & 0.7 & 0.5 (0.1) & Y \\
\enddata\\
\tablenotetext{a}{Submillimeter sources detected with SHARC-II.  For cores with one source detected, the source name is simply the core name from Table \ref{info}.  For cores with multiple sources detected, each source is named individually, and we note which core a source is located in when it is not obvious.  We also include cores with no sources detected and give upper limits for those nondetections.  A ``450'' superscript indicates that the source is located in a core mapped at 450 $\mu$m, otherwise the source is located in a core mapped at 350 $\mu$m.}
\tablenotetext{b}{For detected sources, the value in parentheses indicates the 1$\sigma$ flux uncertainty.  For cores with no sources detected, the value in parentheses indicates the 3$\sigma$ upper limit.  For IRAS 03301+3111, IRAS 03439+3233, EC74, EC88, and L1152-2, we do not present flux densities in 40\arcsec\ apertures because the emission is compact enough to be fully contained in 20\arcsec\ apertures.  For B35A-SMM2, B35A-SMM3, L328-SMM2, and L328-SMM3, we do not present flux densities in 40\arcsec\ apertures because there are nearby sources that overlap in apertures of this size.  For IRAS 04191+1523, we do not present a flux density in a 40\arcsec\ aperture because the source is located too close to the edge of the map.}
\tablenotetext{c}{Distance between peak and Barycenter positions}
\tablenotetext{d}{A Y indicates that \emph{Spitzer} detects a candidate Young Stellar Object within 10\as\ of the submillimeter source, while a N indicates that there is no such \emph{Spitzer} source (see \S\ \ref{embedded}).  A U indicates that \emph{Spitzer} data are unavailable for that source.}
\tablenotetext{e}{There is no candidate YSO detected by \emph{Spitzer} within 10\as, but there is one 11\farcs9 away.  The 350 $\mu$m map suggests L1455-IRS2 may in fact be comprised of two separate submillimeter sources, and although they are not well-resolved enough to present them as two separate sources, the \emph{Spitzer} source is coincident with the western sub-structure.  See text for further discussion.}
\tablenotetext{f}{HH211-MM is associated with an embedded object detected by \emph{Spitzer} only at 70 $\mu$m, thus a Y is present even though it is not detected at wavelengths shortward than and including 24 $\mu$m, and thus not classified as a candidate YSO as defined in \S\ \ref{embedded}.}
\tablenotetext{g}{Confirmed VeLLO}
\tablenotetext{h}{Covered in the map of IRAM 04191+1522}
\tablenotetext{i}{Covered in the map of B18-4}
\tablenotetext{j}{Candidate VeLLO}
\tablenotetext{k}{The map of DC255.4-3.9 covers the positions of three objects classified as candidate VeLLOs based on \emph{Spitzer} data.  The lack of a 350 \um\ detection suggests these candidates are not truly embedded objects (see discussion in \S\ \ref{starless}).}
\tablenotetext{l}{There is an object detected by \emph{Spitzer}, located $\sim$ 9\as\ from L43-SMM3, that has 4.5 and 8.0 $\mu$m fluxes consistent with being a candidate YSO based on the criteria given in \S\ \ref{embedded}.  However, it is not detected by \emph{Spitzer} at 24 or 70 $\mu$m, thus it is unlikely to be an embedded YSO.}
\tablenotetext{m}{The SHARC-II map does not cover the entire B59 core; the observed 350 $\mu$m peak is most likely associated with B59-MMS1 (Reipurth et al. 1996; Brooke et al. 2006)}
\tablenotetext{n}{There are \emph{Spitzer} sources within 10\as\ of each source that appear to be embedded protostars even though they do not meet the criteria for candidate YSOs as described in the text (see text for discussion).}

\end{deluxetable}
\begin{deluxetable}{lcccccccc}
\tabletypesize{\scriptsize}
\tablewidth{0pt}
\tablecaption{\label{properties2}Source Properties derived from Nyquist binned maps}
\tablehead{
\multicolumn{1}{l}{SOURCE\tablenotemark{a}} &\multicolumn{4}{c}{FWHM}&\multicolumn{4}{c}
{2$\sigma$ level}\\
\colhead{}              &
\colhead{Major axis}          &
\colhead{Minor axis}          &
\colhead{Aspect}              &
\colhead{Position}            &
\colhead{Major axis}          &
\colhead{Minor axis}          &
\colhead{Aspect}              &
\colhead{Position}            \\
\colhead{}                    &
\colhead{($\arcsec$)}         &
\colhead{($\arcsec$)}         &
\colhead{ratio}               &
\colhead{angle}               &
\colhead{($\arcsec$)}         &
\colhead{($\arcsec$)}         &
\colhead{ratio}               &
\colhead{angle}
}               
\startdata
L1455-IRS5      & 25 & 18 & 1.35 &  -45.5 &  53 &   41 &   1.27 &   71.1 \\
L1455-IRS1      & 10 & 10 & 1.01 &  -46.3 &  32 &   26 &   1.24 &   71.8 \\
L1455-IRS4      & 15 & 11 & 1.35 &  -42.8 &  28 &   19 &   1.49 &  -43.9 \\
L1455-IRS2      & 22 & 14 & 1.60 &  -21.0 &  28 &   20 &   1.37 &  -36.8 \\
IRAM 03282+3035 & 11 & 10 & 1.02 &  -47.0 &  31 &   24 &   1.29 &  -69.8 \\
IRAS 03292+3039 & 11 & 10 & 1.01 &  -58.5 &  34 &   29 &   1.15 &  -18.6 \\
BOLO68          & 19 & 14 & 1.36 &  -72.5 &  45 &   25 &   1.80 &  -71.1 \\
HH211-MM        & 11 & 10 & 1.03 &  -47.9 &  47 &   33 &   1.41 &  -47.6 \\
IRAS 03445+3242 & 15 & 13 & 1.11 &   79.1 &  39 &   29 &   1.37 &   83.2 \\
IRAM 04191+1522 & 19 & 14 & 1.36 &   67.0 &  32 &   24 &   1.31 &   45.7 \\
IRAS 04191+1523 & -  & -  & -    &   -    &  30 &   19 &   1.56 &   48.9 \\
L1521F          & 16 & 16 & 1.05 &   33.5 &  42 &   28 &   1.48 &   52.0 \\
B18-4           & 27 & 19 & 1.40 &  -28.5 &  31 &   21 &   1.46 &  -14.1 \\
IRAS 04325+2402 & 19 & 13 & 1.38 &  -67.6 &  35 &   33 &   1.06 &    2.2 \\
L1582B          & 11 & 10 & 1.05 &   61.3 &  28 &   26 &   1.10 &  -53.1 \\
B35A-SMM1       & 17 & 14 & 1.22 &   39.3 &  59 &   29 &   2.06 &  -13.5 \\
CG30-SMM1       & 10 & 10 & 1.00 &   40.8 &  -  &   -  &   -    &   -    \\ 
CG30-SMM2       & 11 & 10 & 1.01 &   54.7 &  -  &   -  &   -    &   -    \\
L43-RNO91       & 21 & 18 & 1.20 &  -48.7 &  45 &   31 &   1.42 &  -20.6 \\
L43-SMM1        & 10 & 10 & 1.01 &   46.3 &  -  &   -  &   -    &   -    \\ 
L43-SMM2        & 35 & 19 & 1.80 &   44.6 &  52 &   30 &   1.74 &   42.7 \\
L146            & 17 & 11 & 1.49 &   29.3 &  35 &   32 &   1.10 &  -27.8 \\
B59             & 11 & 7  & 1.50 &  -48.1 &  33 &   27 &   1.22 &   87.2 \\
L507            & 19 & 9  & 2.17 &  -71.1 &  47 &   32 &   1.45 &  -66.5 \\
L328-SMM1       & 32 & 12 & 2.56 &  -70.1 &  50 &   39 &   1.27 &  -65.2 \\
L483            & 24 & 14 & 1.68 &  -14.4 &  67 &   33 &   1.98 &  -12.4 \\
L723            & 15 & 9  & 1.58 &  -76.4 &  35 &   30 &   1.19 &  -89.0 \\
CB188           & 17 & 8  & 1.95 &   45.5 &  22 &   16 &   1.36 &   62.1 \\
B335            & 11 & 10 & 1.02 &  -41.1 &  29 &   27 &   1.09 &  -76.0 \\
L1152-1         & 12 & 11 & 1.07 &  -87.8 &  80 &   41 &   1.94 &  -34.4 \\
L1152-2         & 10 & 5  & 1.95 &    0.0 &  14 &   10 &   1.47 &   69.6 \\
L1041-2         & 21 & 9  & 2.11 &   -9.7 &  39 &   31 &   1.27 &  -35.4 \\
L1157           & 16 & 10 & 1.54 &    1.2 &  47 &   29 &   1.60 &    6.6 \\
L1228           & 11 & 10 & 1.12 &   46.3 &  22 &   17 &   1.30 &  -66.5 \\
Bern48          & 11 & 10 & 1.02 &  -73.5 &  25 &   21 &   1.18 &  -76.4 \\
L1177           & 15 & 9  & 1.53 &   70.7 &  34 &   29 &   1.17 &    8.1 \\
L1014           & 18 & 16 & 1.11 &   62.8 &  32 &   23 &   1.38 &  -66.4 \\
L1165           & 16 & 10 & 1.50 &   86.0 &  46 &   32 &   1.42 &  -86.8 \\
L1221-SMM1      & 10 & 5  & 2.00 &   90.0 &  46 &   33 &   1.41 &   89.6 \\
L1221-SMM2      & 11 & 10 & 1.02 &  -51.4 &  28 &   19 &   1.46 &  -24.6 \\
L1251A-1        & 12 & 12 & 1.01 &  -84.8 &  42 &   28 &   1.50 &  -23.9 \\
L1251A-2        & 18 & 11 & 1.68 &   44.2 &  51 &   20 &   2.50 &   39.4 \\
L1251C          & 6  & 5  & 1.14 &  -83.0 &  20 &   14 &   1.39 &   21.5 \\
L1251B-SMM1     & 24 & 19 & 1.29 &   24.8 &  55 &   35 &   1.59 &   19.7 \\
\hline
L507$^{450}$    & 19 & 11 & 1.73 &  -56.9 &  31 &   18 &   1.67 &  -70.4 \\
L483$^{450}$    & 24 & 14 & 1.68 &  -14.4 &  67 &   33 &   1.98 &  -12.4 \\
L673-7$^{450}$  & 15 & 13 & 1.08 &   76.2 &  32 &   22 &   1.45 &   -7.5 \\
\enddata\\
\tablenotetext{a}{A source with superscript ``450'' indicates that the source is located in a core mapped at 450 $\mu$m.}
\end{deluxetable}
\begin{table}
\caption{\label{l1544model}Observations and Model of L1544}  
\begin{tabular}{lccr}
\hline
Wavelength	& Aperture		& $S_{\nu}^{mod}$	& $S_{\nu}^{obs}$	\\
(\um)		& (\as)			& (Jy)			& (Jy)			\\
\hline
90		& 150			& 0.09			& $<$ 0.57$^a$		\\
170		& 150			& 8.64			& 14.9 (4.5)$^a$	\\
200		& 150			& 15.1			& 18.6 (5.6)$^a$	\\
350		& 40			& 4.5			& $<$ 0.05$^b$		\\
450		& 40			& 4.0			& 4.2 (0.8)$^c$		\\
850		& 40			& 1.31			& 1.12 (0.07)$^c$	\\
1300		& 40			& 0.45			& 0.27 (0.04)$^c$	\\
\hline
\end{tabular}\\
$^a$Ward-Thompson et al. (2002)\\
$^b$This survey\\
$^c$Shirley et al. (2000)\\
\end{table}


\begin{thebibliography}{}

\bibitem[Bertin (2003)]{SExtractor}Bertin, E., SExtractor v2.3 User's Manual (Paris: Insitut d'Astrophysique)

\bibitem[Beelen et al. (2006)]{2006ApJ...642...694}Beelen, A., Cox, P., Benford, D.J., Dowell, C.D., Kovacs, A., Bertoldi, F., Omont, A., \& Carilli, C. 2006, ApJ, 642, 694

\bibitem[Bence et al. (1998)]{1998MNRAS.299..965B}Bence, S.J., Padman, R., Isaak, K.G., Wiedner, M.C., \& Wright, G.S. 1998, MNRAS, 299, 965

\bibitem[Brooke (2006)]{brooke...inpress}Brooke, T. et al. 2006, ApJ, in press

\bibitem[Bourke et al. (2006)]{2006ApJ...649...L37}Bourke, T.L., et al. 2006, ApJ, 649, L37

\bibitem[Caselli et al. (2002)]{2002ApJ...572...238}Caselli, P., Benson, P.\ J., Myers, P.\ C., \& Tafalla, M. 2002, \apj, 572, 238

\bibitem[Crapsi et al. (2005)]{2005ApJ...619...379}Crapsi, A., Caselli, P., Walmsley, C.M., Myers, P.C., Tafalla, M., Lee, C.W., \& Bourke, T.L. 2005, ApJ, 619, 379

\bibitem[de Geus et al.(1989)]{1989A&A...216...44D} de Geus, E.~J., de 
Zeeuw, P.~T., \& Lub, J.\ 1989, \aap, 216, 44

\bibitem[Di Francesco et al. (2006)]{2006PPV}Di Francesco, J.\ D., Evans N.\ J\ .II, Caselli, P., Myers, P. C., Shirley, Y.\ L.,  Aikawa, A., and Tafalla, M. 2006, in Protostars and Planets V, ed. B. Reipurth, D. Jewitt, \& K. Keil (Tucson: Univ. Arizona Press), in press

\bibitem[Dobashi et al. (1994)]{1994ApJS...95..419D} Dobashi, K., Bernard, 
J.-P., Yonekura, Y., \& Fukui, Y.\ 1994, \apjs, 95, 419

\bibitem[Dowell et al. (2002)]{2002SPIE...4855}Dowell, C.D., et al. 2002, SPIE, 4855

\bibitem[Dunham et al. (2006)]{2006ApJ...sub2}Dunham, M.M., et al. 2006, ApJ, 651, 945

\bibitem[Enoch et al. (2006)]{2006ApJ...638...293}Enoch, M.L., et al. 2006, ApJ, 638, 293

\bibitem[Evans et al. (2005)]{datadelivery3}Evans, N. J., II et al. 2005,
``Third Delivery of Data from the c2d Legacy Project: IRAC and MIPS'' (Pasadena, SSC), http://ssc.spitzer.caltech.edu/legacy/

\bibitem[Evans et al. (2003)]{2003PASP...115...965}Evans, N.J. II et al. 2003, PASP, 115, 965

\bibitem[Evans et al. (2001)]{2001ApJ...557...193}Evans, N.J. II, Rawlings, J.M.C., Shirley, Y.L., \& Mundy, L.G. 2001, ApJ, 557, 193

\bibitem[Fazio et al. (2004)]{2004ApJS...154...10}Fazio, G.G., et al. 2004, ApJS, 154, 10

\bibitem[Franco (1989)]{1989AA...223...313}Franco, G.~A.~P.\ 1989, A\&A, 223, 313

\bibitem[Goldsmith et al. (1984)]{1984ApJ...286..599G} Goldsmith, P.~F., 
Snell, R.~L., Hemeon-Heyer, M., \& Langer, W.~D.\ 1984, \apj, 286, 599

\bibitem[Gueth and Guilloteau (1999)]{1999AA...343...571}Gueth, F., and Guilloteau, S. 1999, A\&A, 343, 571

\bibitem[Hartmann et al. (1999)]{1999AJ.118..1784}Hartmann, L., Calvet, N., Allen, L., Chen, H., \& Jayawardhana, R. 1999, AJ, 118, 1784

\bibitem[Harvey et al. 2006a]{2006ApJ...inpress1}Harvey, P.M. et al. 2006a, ApJ, 644, 307

\bibitem[Herbig \& Jones (1983)]{1983AJ.....88.1040H}Herbig, G.~H., \& 
Jones, B.~F.\ 1983, \aj, 88, 1040 

\bibitem[Herbst (1975)]{1975AJ.....80..212H}Herbst, W.\ 1975, \aj, 80, 212

\bibitem[Hirano et al. (2006)]{2006ApJ...636L.141H}Hirano, N., Liu, S.-Y., Shang, H., Ho, P.T.P., Huang, H.-C., Kuan, Y.-J., McCaughrean, M.J., \& Zhang, Q. 2006, ApJ, 636, L141

\bibitem[Ivezic et al. (1999)]{dustymanual}Ivezic, A., Nenkova, M., \& Elitzur, M. 1999, User Manual for DUSTY (Lexington: Univ. Kentucky)

\bibitem[J\o rgensen et al. (2006)]{2006ApJ...inpress2}J\o rgensen, J.K. et al. 2006, ApJ, 645, 1246

\bibitem[Kawamura et al. (2001)]{2001PASJ...53.1097K} Kawamura, A., Kun, M., 
Onishi, T., Vavrek, R., Domsa, I., Mizuno, A., \& Fukui, Y.\ 2001, \pasj, 
53, 1097

\bibitem[Kenyon et al. (1994)]{1994AJ....108.1872K} Kenyon, S.~J., Dobrzycka, D., \& Hartmann, L.\ 1994, \aj, 108, 1872

\bibitem[Kirk et al. (2006)]{2006ApJ...646...1009} Kirk, H., Johnstone, D., \& Di Francesco, J. 2006, ApJ, 646, 1009

\bibitem[Kun (1998)]{1998ApJS..115...59K} Kun, M.\ 1998, \apjs, 115, 59

\bibitem[Kun \& Prusti (1993)]{1993A&A...272..235K} Kun, M., \& Prusti, T.\ 
1993, \aap, 272, 235 

\bibitem[Lee et al. (2001)]{2001ApJS...136...703}Lee, C\ W., Myers, P.\ C., \& Tafalla, M. 2001, \apjs, 136, 703

\bibitem[McCaughrean et al. 1994]{1994ApJ...436...189}McCaughrean, M.J., Rayner, J.T., \& Zinnecker, H. 1994, ApJ, 436, 189

\bibitem[Motte \& \andre\ (2006)]{2001AA...365...440}Motte, F., \& \andre\, P. 2001, A\&A, 365, 440

\bibitem[Murdin et al. (1977)]{1977MNRAS...181...657}Murdin, P. \& Penston, M.V. 1977, MNRAS, 181, 657

\bibitem[oh5]{1994A&A...291...943}Ossenkopf, V., \& Henning, T. 1994, A\&A, 291, 943

\bibitem[Pagani \& Breart de Boisanger (1996)]{1996A&A...312..989P}Pagani, 
L., \& Breart de Boisanger, C.\ 1996, \aap, 312, 989

\bibitem[Palau et al. (2006)]{2006ApJ...636L.137P}Palau, A., Ho, P. T. P., Zhang, Q., Estalella, R., Hirano, N., Shang, H., Lee, C.-F., Bourke, T. L., Beuther, H., \& Kuan, Y.-J. 2006, ApJ, 636, L137

\bibitem[Porras et al. 2006]{porrasinprep}Porras, A. et al. 2006, ApJ, in press

\bibitem[rebull2006]{2006rebull}Rebull, L.M. et al. 2006, submitted to ApJ

\bibitem[Rieke et al. (2004)]{2004ApJS...154...25}Rieke, G.H., et al. 2004, ApJS, 154, 25

\bibitem[Shirley et al. (2002)]{2002ApJ...575...337}Shirley, Y.\L., Evans, N.\ J.\ II, and Rawling, J.\ M.\ C. 2002, ApJ, 575, 337

\bibitem[Shirley et al. (2000)]{2000ApJS...131...249}Shirley, Y.\ L., Evans, N.\ J.\ II, Rawling, J.\ M.\ C., and Gregersen, E. M. 2000, \apjs, 131, 249

\bibitem[Shu et al. 1987]{1987ARAA...25...23}Shu, F.H., Adams, F.C., \& Lizano, S. 1987, ARA\&A, 25, 23

\bibitem[Strai{\v z}ys et al. (2003)]{2003A&A...405..585S} Strai{\v z}ys, 
V., {\v C}ernis, K., \& Barta{\v s}i{\= u}t{\.e}, S.\ 2003, \aap, 405, 585

\bibitem[Strai{\v z}ys et al. (1996)]{1996BaltA...5...125}Strai{\v z}ys, V., {\v C}ernis, K., \& Barta{\v s}i{\= u}t{\.e}, S.\ 1996, Baltic Astronomy, 5, 125

\bibitem[Straizys et al. (1992)]{1992BaltA...1..149S} Strai{\v z}ys, V., {\v
C}ernis, K., Kazlauskas, A., \& Mei{\v s}tas, E.\ 1992, Baltic Astronomy,
1, 149

\bibitem[Terebey et al. (1993)]{1993ApJ...414...759}Terebey, S., Chandler, C.J., \& \andre, P. 1993, ApJ, 414, 759

\bibitem[Ward-Thompson et al. (2002)]{2002MNRAS.329.257}Ward-Thompson, D., \andre, P., \& Kirk, J.M. 2002, MNRAS, 329, 257

\bibitem[Woermann et al. (2001)]{2001MNRAS.325.1213W} Woermann, B., Gaylard, 
M.~J., \& Otrupcek, R.\ 2001, \mnras, 325, 1213

\bibitem[Yonekura et al. (1997)]{1997ApJS..110...21Y} Yonekura, Y., Dobashi, 
K., Mizuno, A., Ogawa, H., \& Fukui, Y.\ 1997, \apjs, 110, 21

\bibitem[Young et al. (2006)]{inprepc}Young, C.H., et al. 2006, AJ, 132, 1998

\bibitem[Young et al. (2004)]{2004ApJS...154...396}Young, C.H., et al. 2004, \apjs, 154, 396

\bibitem[Young et al. (2003)]{2003ApJS...145...111}Young, C.H., Shirley, Y.L., Evans, N.J. II, \& Rawlings, J.M.C. 2003, ApJS, 145, 111

\end{thebibliography}
\end{document}